\documentclass[twocolumn,tighten,twocolappendix]{aastex631}%linenumbers,,

\usepackage{mathtools}
\usepackage{chngcntr}
\usepackage{amssymb}
\usepackage{amsmath}
\usepackage{color}
\usepackage{rotating,graphicx}
\def\hin{{{\rm H}\,{\sc i}}}
\def\ov{{{\rm O}\,{\sc v}~}}
\def\ovi{{{\rm O}\,{\sc vi}~}}
\def\ovin{{{\rm O}\,{\sc vi}}}
\def\ovii{{{\rm O}\,{\sc vii}~}}
\def\oviin{{{\rm O}\,{\sc vii}}}
\def\oviii{{{\rm O}\,{\sc viii}~}}

\def\planck{{\it Planck}~}
\def\planckn{{\it Planck}}
\def\act{{\it Atacama Cosmology Telescope}~}
\def\actn{{\it Atacama Cosmology Telescope}}
\def\jvla{{\it Jansky Very Large Array}}
\def\spt{{\it South Pole Telescope}~}
\def\msun{{M$_{\odot}$}~}
\def\msunn{{M$_{\odot}$}}
\def\rvirial{{R$_{200}$}~}
\def\rvirialn{{R$_{200}$}}

\defcitealias{Das2023a}{D23}
\defcitealias{Madhavacheril2020}{M20}
\defcitealias{Zou2019}{Z19}
\defcitealias{Arnaud2010}{A10}
\defcitealias{Kim2022}{K22}
\defcitealias{Amodeo2021}{A21}

\shorttitle{tSZ effect and star formation}
\shortauthors{Das et al.}

\begin{document}

\title{Thermal Sunyaev-Zel'dovich Effect in the circumgalactic medium - II: dependence on star formation}

\correspondingauthor{Sanskriti Das}
\email{snskriti@stanford.edu; dassanskriti@gmail.com}

%\author[0000-0002-9069-7061]{Sanskriti Das (\textbengali{সংস্কৃতি দাস)}}
\author[0000-0002-9069-7061]{Sanskriti Das}
\altaffiliation{Hubble Fellow}
\affil{Kavli Institute for Particle Astrophysics and Cosmology, Stanford University, 452 Lomita Mall, Stanford, CA\,94305, USA}

\author[0000-0003-4983-0462]{Nhut Truong}
\affil{Center for Space Sciences and Technology, University of Maryland, Baltimore County, 1000 Hilltop Circle, Baltimore, MD\,21250, USA}
\affil{Goddard Space Flight Center, Greenbelt, MD\,20771, USA}

\author[0000-0001-6320-261X]{Yi-Kuan Chiang}
\affiliation{Academia Sinica Institute of Astronomy and Astrophysics (ASIAA), No. 1, Section 4, Roosevelt Road, Taipei 10617, Taiwan}

\author[0000-0002-4822-3559]{Smita Mathur}
\affil{Department of Astronomy, The Ohio State University, 140 W. 18th Ave., Columbus, OH 43210, USA}
\affil{Center for Cosmology and AstroParticle Physics, The Ohio State University, 191 W. Woodruff Ave., Columbus, OH 43210, USA}

%\affiliation{Institute of Astronomy and Astrophysics, Academia Sinica, Astronomy-Mathematics Building, No.\,1, Section\,4, Roosevelt Road, Taipei 10617, Taiwan}

\begin{abstract}
We measure thermal Sunyaev{\textendash}Zel{\textquoteright}dovich (tSZ) Effect in the circumgalactic medium (CGM) of $\approx$2.5 million $\rm M_{200}$=$\rm 10^{12-14}$\msun WISE$\times$DESI galaxies out to $z$=1.2. We split the sample into quiescent (0.7 million) and star-forming (1.8 million) galaxies, exploring the relation between the thermal pressure of the CGM and star formation for the first time. We develop and implement a novel probabilistic approach to cross-correlate the galaxy catalog with the \actn$+$\planck data by taking into account the uncertainties in redshift, mass, and star formation rate. The S/N of the stacked Compton-$y$ value in the CGM varies from 4.9 to 18.5, depending on the sample size and the CGM signal strength within the relevant mass bin. We detect the CGM signal down to $\rm M_{200}=10^{12.3}$\msunn, and provide stringent upper limit at $\rm M_{200}<10^{12.3}$\msunn. The data fit well with the standard GNFW profile of thermal pressure and do not require a flatter or steeper profile. This suggests a significant impact of cooling and the absence of dominant feedback. In galaxies with $\rm M_{200}\approx10^{12.3-12.8}$\msun halos, the volume filling CGM is likely the largest contributor to the \textit{galactic} baryons at their virial temperatures of $\sim$10$^{6-6.4}$\,K. For $\rm M_{200}>10^{12.8}$\msun halos, the most massive phase of the CGM is likely at a sub-virial temperature of $\gtrsim 10^6$\,K. The thermal energy of the CGM of quiescent galaxies follows the self-similar relation with mass, but the star-forming galaxies deviate from this relation. This indicates that the impact of non-gravitational factors varies among halos of different degrees of star-forming activity.
\end{abstract}

\keywords{Sunyaev-Zeldovich effect --- Millimeter astronomy --- Extragalactic astronomy --- Observational Cosmology --- Circumgalactic medium --- Hot Intergalactic medium --- Hot ionized medium --- Galaxy evolution --- Galaxy environments --- Galactic winds --- Galaxy processes --- Cosmic microwave background radiation}

\section{Introduction}\label{sec:intro}

The most massive and volume-filling phase of the multiphase circumgalactic medium (CGM) is predicted to be $\geqslant 10^6$\,K hot and fully ionized \citep[e.g.,][]{Spitzer1956,Schaye2015,Mathur2022,Truong2023}. If galaxy evolution is primarily driven by gravity, the physical properties of the CGM would solely depend on the virial mass, referred to as self-similarity. However, it is predicted that the self-similarity breaks in the CGM (of lower mass halos) due to the increasingly important role of galactic feedback compared to the gravity \citep[e.g.,][]{Lim2021,Kim2022,Pop2022}.

%SM
The Sunyaev-Zel'dovich (SZ) effect \citep{Sunyaev1969,Mroczkowski2019} is a powerful technique to characterize the hot CGM because of its redshift and metallicity independence, and linear density dependence. The SZ effect is a distortion in the cosmic microwave background (CMB) spectrum due to the inverse Compton scattering of the CMB photons with the free electrons in the intervening plasma. Thermal SZ (tSZ) effect, the strongest of different SZ effects, is characterized by the Compton-$y$ parameter: $y(r_\perp) = (\sigma_{\rm T}/m_e c^2) \int {\rm P_e(r)} dr_\parallel$\footnote{$\sigma_{\rm T}, m_e$ and c are Thompson's scattering cross section, rest mass of the electron, and the speed of light in vacuum, respectively.}. It is a measure of the thermal pressure (${\rm P_e} = n_e \rm k_B T_e$) or thermal energy ($\propto \int{\rm P_e dV} \propto \int y \rm dA$) of the free electrons of the relevant medium integrated along the line-of-sight. %the thermal energy of the halo gas ($\propto \int{\rm P_e dV} \propto \int y \rm dA$)

In the absence of any dominant feedback, the thermal pressure of the CGM would primarily depend on the virial mass. In that case, star-forming and quiescent galaxies of the same mass would have a similar tSZ effect. If, however, the galactic scale stellar feedback plays the dominant role, the CGM of star-forming galaxies would have a stronger tSZ effect. On the other hand, if the past AGN feedback plays a dominant role, the CGM of a quiescent galaxy would show a stronger tSZ effect. Thus, studying the tSZ effect as a function of star-forming activities is imperative.

\begin{figure*}
    \centering
    \includegraphics[width=0.995\linewidth]{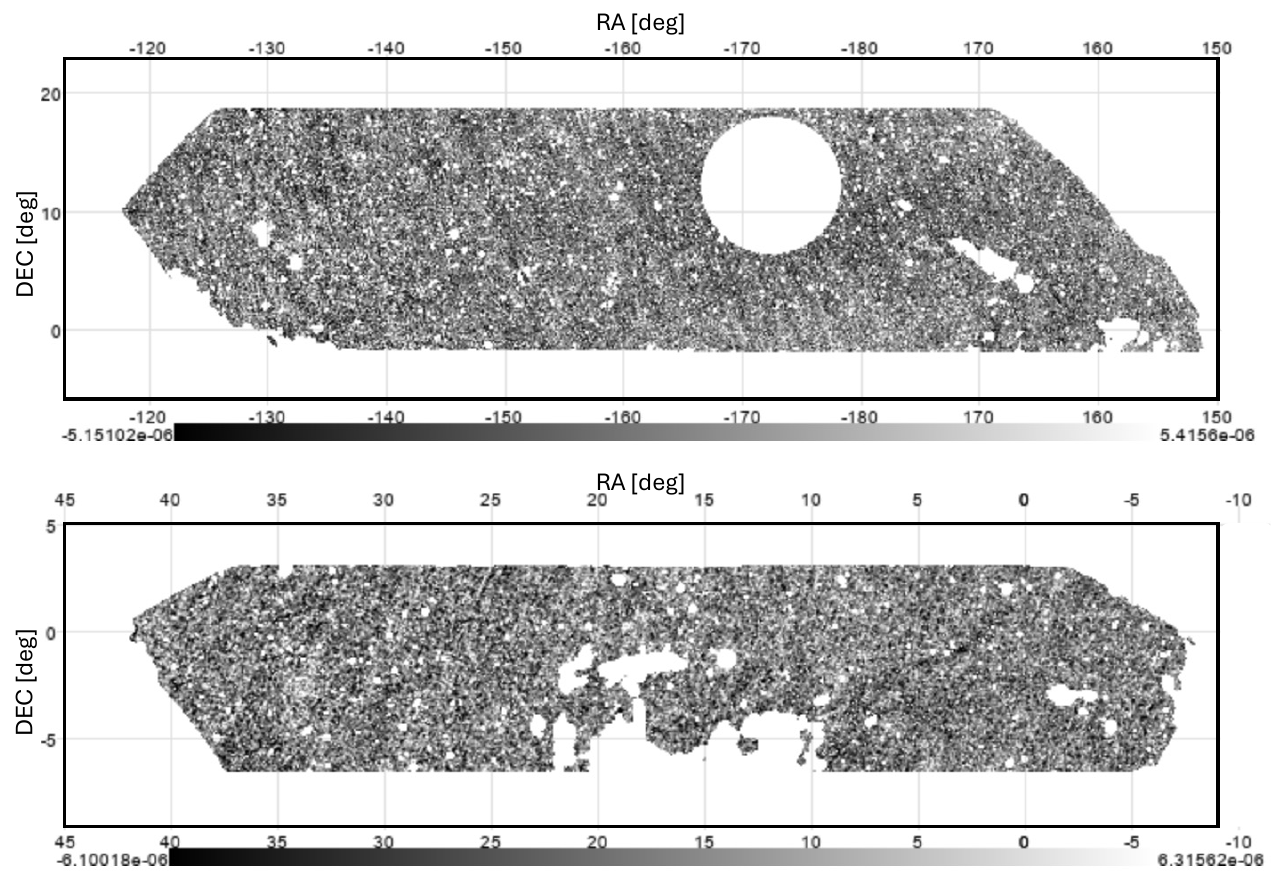}
    %%\vspace{-0.1 in}
    \caption{CIB-deprojected Compton-$y$ maps (top: \texttt{BN}, bottom: \texttt{D56}) from \citetalias{Madhavacheril2020} used in our analysis, with regions around high Galactic dust extinction, SZ-detected galaxy clusters, and radio-loud sources masked.}
    \label{fig:ymaps}
\end{figure*}

\begin{figure*}
    \centering
\includegraphics[width=0.497\textwidth]{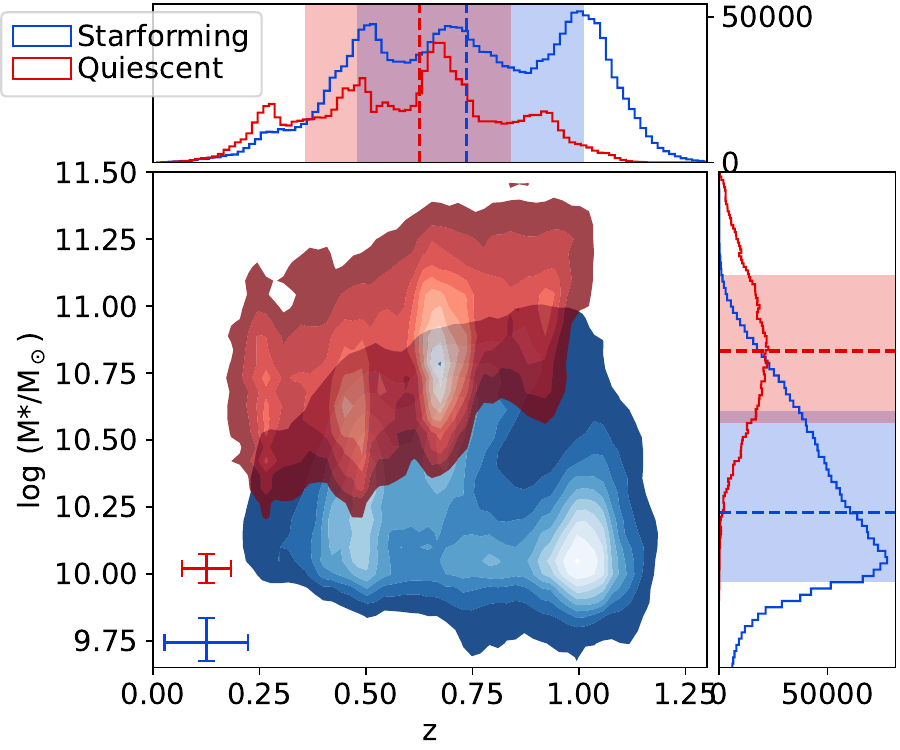}
\includegraphics[width=0.497\textwidth]{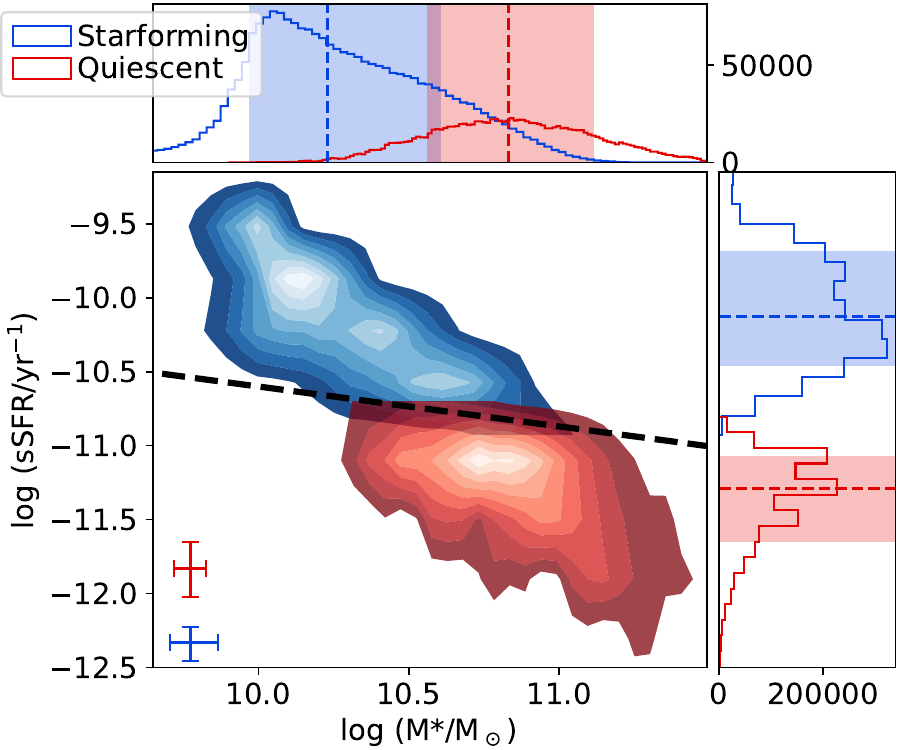}
%\includegraphics[width=0.32\textwidth]{Mvir_z.pdf}
    %%\vspace{-0.3 in}
    \caption{Distribution of stellar mass with redshift (left panel) and sSFR (right panel) of our galaxy sample from \citetalias{Zou2019}. The blue and red histograms/contours correspond to the star-forming and quiescent galaxies, respectively. The dashed lines and shaded areas around them denote the median and 68\% confidence interval of the relevant parameter. %after excluding potential group or cluster members, AGN, and galaxies with poorly constrained mass and SFR. 
    Average uncertainties in stellar mass, sSFR (=SFR/$\rm M_\star$), and redshift are shown in the bottom left corner of the plots; these parameters are generally better constrained at higher mass. The black dashed line in the right panel denotes the median star-forming main sequence from \cite{Bluck2016}.}% The grid in the right panel corresponds to the mass and redshift \textit{slices} considered in the stacking (see \S\ref{sec:stack}). Note that the redshift and mass of some galaxies lie beyond our considered range, but they have a non-negligible probability of being within the range because of large uncertainties in redshift and mass.}
    \label{fig:Mz}
\end{figure*}

Due to the large beam size of \planck (FWHM $\approx 10'$), studies of the tSZ effect were initially limited to the CGM of M$_\star$$>$10$^{11.3}$ \msun locally brightest galaxies (LBGs) and galaxy groups at $z\lesssim 0.1$ \citep{Planck2013,Greco2015,Lim2018} and the large-scale structure of the universe up to $z$=1 \citep{Chiang2020,Chiang2021}. The better sensitivity and smaller beamsize (FWHM $\approx 1.\!'4$) of \act (ACT) and \spt (SPT) than \planck have enabled resolving smaller halos than before. The combined maps of ACT{+}\planck and SPT{+}\planck have been cross-correlated with catalogs of massive galaxies and galaxy groups at $z \leqslant 1.5$ \citep{Schaan2021,Meinke2021,Vavagiakis2021,Pandey2022,Sanchez2023,Liu2025}. In \cite{Das2023a}, hereafter \citetalias{Das2023a}, we extended this study to lower masses, focusing on 0.6 million $z\approx$ 0--0.3 WISE$\times$SuperCosMos galaxies using \planck\!{+}ACT data, the largest galaxy sample studied in the tSZ effect at that time. We detected the tSZ effect in M$_\star$= 10$^{10.6-11.3}$ \msun galaxies and provided the stringent upper limit for the M$_\star$=10$^{9.8-10.6}$ \msun galaxies, establishing that the tSZ effect in lower mass halos can be constrained. 

However, the majority, if not the entirety, of existing studies conducted so far have examined halos through a cosmological lens, treating virial mass as the sole descriptor of halos. Also, massive halos have been commonly targeted because of their larger sizes and more pronounced tSZ effect. Because the central galaxies in these massive halos often have multi-wavelength observational evidence of AGN feedback independent from SZ, any observed deviations from a gravity-dominated regime are usually attributed to AGN feedback. In contrast, in lower-mass halos, the complexity of diverse feedback and cooling mechanisms may render a simplistic gravity vs feedback model inadequate. Therefore, adopting a complementary astrophysical perspective is essential. As the initial step toward this new paradigm, we expand the analysis to 2.5 million WISE$\times$DESI galaxies of comparable mass to \citetalias{Das2023a} over a wider redshift range, facilitating the first investigation into any correlation between star-forming activities and thermal pressure in the ionized CGM. This is to date the largest galaxy sample explored in the tSZ effect \textcolor{black}{cross-correlating $y$-maps and galaxy catalogs}.

We have used the flat $\Lambda$CDM (cold dark-matter) cosmology of \cite{Planck2020}: local expansion rate $ H_0 =67.66\rm\, km\,s^{-1} Mpc^{-1}$, matter density $\Omega_m=0.30966$, baryon density $\Omega_b=0.04897$, and dark energy density $\Omega_\Lambda = 1- \Omega_m = 0.69034$ throughout the paper. The cosmological baryon fraction $f_{b,cosmo} = \Omega_b/\Omega_m$ is 0.15814. Virial parameters are expressed in terms of over-density $\Delta$, e.g., ${\rm M_\Delta = \frac{4}{3} \pi R_\Delta^3 }\Delta\rho_c(z)$ is the mass enclosed within a sphere of radius $\rm R_\Delta$, where the mean mass density is $\Delta$ times the critical density of the universe, $\rho_c(z) = 3H(z)^2/8\pi \rm G$. $H(z)$ is the Hubble parameter at redshift $z$, and G is the Newtonian constant of gravitation. The redshift evolution of the Hubble parameter is $E(z)^2 = H(z)^2/H_o^2 = \Omega_m\times(1+z)^3 + \Omega_\Lambda$. 

The paper is organized as follows. In \S\ref{sec:data}, we discuss the details about the data and analysis methodology. We present and interpret our results in \S\ref{sec:result}. We summarize our results and discuss future directions in \S\ref{sec:end}. 

\section{Data extraction and analysis}\label{sec:data}

Here we provide the details about the Compton-$y$ map (\S\ref{sec:ymap}) and the galaxy sample (\S\ref{sec:galsample}) to cross-correlate with, methods to stack $y$-maps and obtain radial profiles of the thermal pressure from it (\S\ref{sec:stack}), and extraction of the CGM signal through component separation of the thermal pressure (\S\ref{sec:P}). We derive the thermal energy and baryon fraction from the extracted CGM signal in \S\ref{sec:Y} and \S\ref{sec:masscal}, respectively.

\subsection{Compton-$y$ maps}\label{sec:ymap}

We consider the Compton-$y$ maps constructed from ACTPol\,DR4 and \planck data in two non-overlapping patches of the sky: BOSS (Baryon Oscillation Spectroscopic Survey) North or \texttt{BN} ($-117^\circ < \rm RA<150^\circ$, $-2^\circ < \rm DEC < 19^\circ$; area = 1633 deg$^2$) and Deep\,56 or \texttt{D56} ($-9^\circ < \rm RA<40^\circ$, $-7^\circ < \rm DEC < 4^\circ$; area = 456 deg$^2$) from \citet[hereafter \citetalias{Madhavacheril2020}]{Madhavacheril2020}. In further analyses, we use the dust-corrected $y$-maps masked in the position of galaxy clusters and radio sources (see Figure\,\ref{fig:ymaps}). Below, we summarize the procedure discussed in detail in \S2 of \citetalias{Das2023a}. 

To account for the effect of thermal dust, we consider the Compton-$y$ maps from \citetalias{Madhavacheril2020} after the deprojection of the cosmic infrared background (CIB). To minimize the effect of Galactic dust, we exclude the regions corresponding to neutral hydrogen column density of the Galaxy, N(\hin), above 4.7$\times 10^{20} \rm cm^{-2}$ and $4.2\times 10^{20} \rm cm^{-2}$ for the sky regions \texttt{BN} and \texttt{D56}. The average N(\hin) in these regions after masking is $2.4_{-0.8}^{+1.0}\times 10^{20} \rm cm^{-2}$ and $2.9\pm0.6\times 10^{20} \rm cm^{-2}$, respectively.

We mask the $\theta_{\rm 200,cl} =  \rm R_{200, cl}/D_{\rm A,cl}$\footnote{$\rm R_{ 200}$ is the virial radius, and D$\rm _A$ is the angular diameter distance.} regions around all galaxy clusters in the tSZ galaxy cluster catalog \citep{Hilton2021}, the Virgo cluster, and Abell\,119 cluster to avoid the contamination by the tSZ effect of the intracluster medium (ICM). Using the latest catalog of the FIRST \citep[Faint Images of the Radio Sky at Twenty-cm;][]{FIRST1997} survey, we also mask the regions affected by the radio sources down to the 1$\sigma$ noise level of ACT at 98 and 150\,GHz to eliminate the contamination by the emission from compact radio sources. 

%We ``in-paint" the smaller masked regions, i.e., galaxy clusters in the catalog of \cite{Hilton2021} and radio sources in \cite{FIRST1997} with biharmonic equations \citep{scikitimage2014}. To avoid artifacts created by the ``in-paint", larger regions masked by the Galactic dust and Virgo cluster are kept masked and ignored in the following analysis. 

\subsection{Galaxy sample}\label{sec:galsample}

To cross-correlate with the Compton-$y$ maps, we consider the \textcolor{black}{photometric redshifts, stellar mass, and star formation rates (SFR) from the} galaxy catalog of \cite{Zou2019}, hereafter \citetalias{Zou2019}. It is based on the optical ($grz$) data from legacy imaging surveys of DESI (Dark Energy Spectroscopic Instrument), and near-infrared (W1 - 3.4$\mu$m, W2 - 4.6$\mu$m) data from \textit{WISE} (Wide-field Infrared Survey Explorer). The photometric spectral energy distribution (SED) of a galaxy is defined by four colors and one magnitude: $g-r,r,r-z,r-W1$, and $r-W2$. The photometric redshift of the galaxy is determined from a local linear regression relation between spectroscopic redshifts and photometric SEDs for $k$-nearest neighbors ($k$ = 150) of the said galaxy in the five-dimensional SED space. The stellar mass and SFR are calculated by fitting the observed photometric SED with stellar population synthesis models, assuming an exponentially declining star formation history with nine values of star formation timescale ranging from 0.1 to 30 Gyr. 

To distinguish galaxies based on their star formation, we consider the median star-forming main sequence, $\rm SFR_{MS}$ \citep{Bluck2016}. 
\begin{equation}\label{eq:sfms}
        \rm log(SFR_{MS}) = 0.73log(M_\star) - 7.3
\end{equation}
Galaxies with $\rm \Delta SFR = log_{10}(SFR/SFR_{MS}) > -0.6$ are star-forming galaxies and the rest of the galaxies are quiescent (Figure\,\ref{fig:Mz}, right).% (which include green valley galaxies). 

We calculate the virial masses of the galaxies, $\rm M_{200,gal}$, from the stellar masses using the stellar-to-halo mass relation (SHMR) of \cite{Bilicki2021}, where the halo masses were calculated using the weak gravitational lensing signal around the galaxies, and stellar masses were calculated from the magnitude in nine ($ugri$ZYJHK$_s$) photometric bands. At the same stellar mass, quiescent galaxies occupy more massive halos than those occupied by star-forming galaxies. We calculate the virial radii, $\rm R_{200,gal}$, using the Navarro-Frank-White (NFW) profile of the dark-matter \citep{Navarro1997}, and determine the corresponding angular size, $\theta_{\rm 200,gal}$ ($\rm = R_{200,gal}/D_{A,gal}$). %Using the concentration factor, $c_{\rm 200}$ as a function of the virial mass \citep{Neto2007}, we convert M$_{\rm 200,gal}$ to M$_{\rm 500,gal}$ and obtain the corresponding R$_{\rm 500,gal}$.

Depending on the strength and mode (thermal or kinetic), the AGN feedback can boost or suppress the CGM signal, and this effect might dominate over the integrated signal from the CGM of stacked galaxies. Interpreting the stacked properties of the CGM would be non-trivial if a non-negligible fraction of galaxies are active. We exclude galaxies with W1--W2$>$0.8 ($\approx$2.4\% of the galaxy sample) so that the remaining galaxies are unlikely to host active nuclei\footnote{Our galaxy sample could have still passed through many cycles of past AGN feedback. We aim to see whether the collective effect of that feedback is detectable. Thus, while in active galaxies, one can assess the current \textit{weather} of the CGM, our goal is to estimate the global \textit{climate} of the ionized CGM.}. By cross-matching the galaxy catalog with the FIRST catalog, we find that $\approx$0.1\% of the galaxy sample are radio-loud; we exclude them from our galaxy sample\footnote{Radio-loud galaxies are excluded for a different reason than general active galaxies. Radio-loud galaxies exhibit large negative Compton-$y$ values at their core even after CIB deprojection. Excluding them does not affect the SZ statistics due to their small number, but it mitigates unphysical negative bias in the stacked measurements. See more details in \citetalias{Das2023a}.}. 

Because the SZ effect is redshift independent, galaxy clusters that are not hosting our target galaxies but are along the line-of-sight of any target galaxy can still contaminate the SZ effect of the CGM. We exclude galaxies whose halos intersect the halo of any galaxy cluster detected in SZ, i.e., galaxies within $\theta_{\rm 200,gal}+\theta_{\rm 200,cl}$ of galaxy clusters; it is $\approx$6.4\% of the galaxy sample.

%The CGM properties of field galaxies and galaxies in groups/clusters might differ because of the surrounding intragroup/cluster medium and other group/cluster members. It makes interpreting the stacked properties of the CGM nontrivial. To alleviate that problem, we focus on field galaxies in this paper. To separate the field galaxies from the group/cluster members, we consider the catalogs of galaxy groups based on the Two Micron All-Sky Redshift Survey and DESI legacy imaging survey\,DR8 \citep{Lim2017,Yang2021}. These catalogs contain galaxy groups at $0<z\leqslant 1.0$ down to the halo mass of 10$^{11.5}$\msun (with 0.45\,dex/0.2\,dex uncertainty below/above 10$^{13.5}$\msun) and richness of 2 with $>$90\% purity. We cross-match the galaxy catalog with the group catalogs and find that $\approx$35\% star-forming galaxies and $\approx$45\% quiescent galaxies are group/cluster members; we exclude them from our galaxy sample. Because of the large uncertainty in photometric redshift, some field galaxies might have been identified as a group/cluster member. Thus our sample of field galaxies could be incomplete but unbiased.   

%Overall, we excluded 46\% of the galaxy sample because of its association with a galaxy group/cluster and/or nuclear activity. 
The SZ effect is expected to correlate with the SFR-dependent halo mass that we calculate from the stellar mass. Therefore, the mass and SFR must be well-constrained. In the following sections, we consider galaxies with uncertainties in stellar mass, halo mass, and SFR each $\leqslant$0.5 dex.%, and uncertainties in $\theta_{\rm 200,gal}$ smaller than the beamsize of the y-map. Thus we end up with 32\% of the galaxy sample we started with. The redshift distribution of the stellar masses for both star-forming and quiescent galaxies is shown in Figure \ref{fig:Mz}. %We focus on galaxies of stellar mass $9.0 \leqslant \rm log(M_\star/M_\odot) \leqslant 11.3$. The median redshift and the median stellar mass of the star-forming galaxies are 0.69 and 10$^{9.8}$\msun, and for quiescent galaxies, it is 0.55 and 10$^{10.6\pm0.2}$\msun. 

\textcolor{black}{Note that we start with $\approx$11 million galaxies in the overlapping sky regions of \texttt{BN, D56} and $\rm WISE\times DESI$. But after our rigorous target selection (discussed above) and masking of the $y$-map (discussed in \S\ref{sec:ymap}), the effective sample size reduces to 2.5 million.}

\begin{figure}
    \centering
    \includegraphics[width=0.995\linewidth]{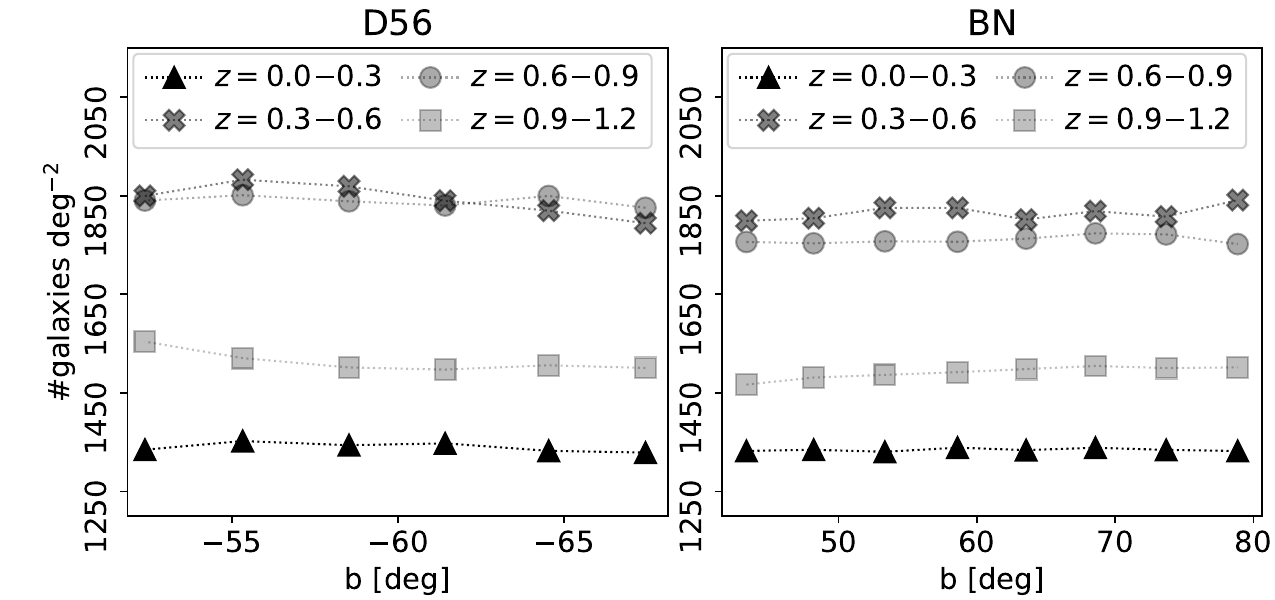}
    %%\vspace{-0.1 in}
    \caption{The surface density distribution of the galaxies as a function of the galactic latitude and the galaxy redshift.}%$9.0 \leqslant \rm log(M_\star/M_\odot) \leqslant 11.3$ 
    \label{fig:surfdens}
\end{figure}

The galaxy catalog \citepalias{Zou2019} is constructed after identifying stars based on their SED and removing them. Moreover, the sky regions \texttt{BN} and \texttt{D56} are unlikely to be contaminated by the stars in the Galactic disk owing to their high galactic latitude. To verify this, we calculate the surface density, i.e., the number of galaxies per unit area of the sky as a function of the galactic latitude and the redshift of the galaxies (Figure\,\ref{fig:surfdens}). In the event of stellar contamination in the galaxy sample, the surface density would be anti-correlated with the galactic latitude. However, the surface density is practically uniform, displaying $<1.5$\% fluctuation across all redshift bins (Figure\,\ref{fig:surfdens}). At a given galactic latitude, the surface density follows the redshift distribution shown in Figure\,\ref{fig:Mz} (left panel), but it bears no correlation with the galactic latitude at a given redshift. This confirms that our galaxy sample is free from contamination by the stars in the Galactic disk.

\subsection{Stacking and aperture photometry}\label{sec:stack}

We develop a novel probabilistic approach in stacking the $y$-stamps by fully accounting for the uncertainties in redshift, SFR, and halo mass, and extract the stacked radial profile of Compton-$y$, $y_{\rm stacked}$, in a computationally rigorous but inexpensive way. The \textcolor{black}{technical} details are provided in appendix\,\ref{sec:stack-technical}.

%We estimate the uncertainties in stacked Compton-$y$ value using the bootstrap method. For a sample of N galaxies, we create a replica of the sample by replacing one galaxy with a randomly chosen galaxy from the rest N-1 galaxies in that sample. For every stack, we make a set of 1000 replicas and obtain the mean and covariance matrix of those replicas.

% redshift bin 0.02 dex, virial mass bin 0.1 dex. Separately done for SF and Q. Same galaxy with different w_SF, M200, and w(M200).  
% Instead of stacking a 2d map that requires reconstruction we extract a 1 d profile and then stack. Faster, less assumption.
%only differential
%battaglia2012 profile add. 

%adaptive mass bin. in virial mass separately for star-forming and quiescent 
%adaptive stamp size so that at least 2 Mpc is probed.
%cubic interpolation
%>50% NAN ignored
%bootstrap replica 1000 or 1% of the sample size whichever is larger. 
%variable zero point handled. cumulative main. differential for cross-checking. 

\subsection{Pressure modeling}\label{sec:P}

$y_{\rm stacked}$ is expected to consist of two components: the ``1-halo" term ($y_{1h}$), i.e., the projected radial profile of Compton-$y$ of stacked galaxies, and the tSZ background. This background consists of the ``2-halo" term, $y_{\rm 2h}$ and any zero-point offset, $y_{zp}$. $y_{2h}$ is the tSZ signal of all other halos around the galaxies being stacked. The $y_{zp}$ comprises the tSZ signal of halos in the foreground/background of stacked galaxies, global Compton-$y$ signal arising from reionization, the intergalactic medium, the CGM of the Milky\,Way, residual contamination from dust in the ISM/CGM of stacked galaxies, and any calibration uncertainties. 

We follow a similar procedure as \citetalias{Das2023a} to model and fit the pressure. We use the generalized NFW (GNFW) profile of thermal pressure of free electrons (Equation\,\ref{eq:pressure}) that was first defined by \cite{Nagai2007} to describe the intracluster medium and employed in \cite{Amodeo2021} for the CGM in a modified way.  %$y_{zp}$ accounts for any residual after null subtraction -- the global and anisotropic Compton-y signal arising from reionization, intergalactic filaments, the CGM of the Milky Way, and uneven calibration uncertainties across the $y$-map. 

%start with the null hypothesis. is 2h enough? is Po, beta and 2h enough? then Po,xc,alpha,beta,2h. then simulations. 

\begin{subequations}
    \begin{equation}\label{eq:pressure}
    P_e(x|M,z) = \frac{\mu}{\mu_e}P_{200}(M,z)P_{\rm GNFW}(x)
\end{equation}
Here $x = r/R_{200}$, $\mu$ is the mean molecular weight, and $\mu_e$ is the mean molecular weight per free electron. $\frac{\mu}{\mu_e}$ converts the gas pressure to free electron pressure. We adopt $\mu$ = 0.59 and $\mu_e$ = 1.14, the values appropriate for a fully ionized medium with primordial abundances.
\begin{equation}
    P_{200}(M,z) = 200\frac{GM_{200}}{2R_{200}}\rho_{c}(z)f_{b,cosmo}
\end{equation}
\begin{equation}\label{eq:pgnfw}
    P_{\rm GNFW}(x) = P_o \Big(\frac{x}{x_c}\Big)^\gamma \Big[ 1+ \Big(\frac{x}{x_c}\Big)^\alpha \Big]^{(\beta-\gamma)/\alpha}
\end{equation}

Here, $x_c$ is the scaled core radius $r_c/R_{200}$, $\alpha$, $\beta$, and $\gamma$ are the slopes of the pressure profile at $x\approx x_c$, $x\gg x_c$, and $x\ll x_c$, respectively. Because of significant degeneracy among the GNFW parameters, we freeze $x_c$, $\alpha$, $\beta$, and $\gamma$, at their empirically derived best-fitted value of 0.6, 1.015, 5.49, and -0.308 for M$_{500}$$>$ 10$^{14}\rm M_\odot$ galaxy clusters at $z<0.2$ \citep{Arnaud2010}. Most halos of our galaxy sample are unresolved by the ACT beam\footnote{The ACT beam is similar to and two times larger than the median \rvirial of quiescent and star-forming galaxies, respectively.}; therefore, the pressure profile is unlikely to be sensitive to the exact values of these parameters. Moreover, if we arbitrarily vary these parameters without specific physical motivation, a flatter pressure profile starts mimicking the shape of the ``2-halo" term to the extent of becoming indistinguishable from each other. To avoid that degeneracy, we keep the shape of the pressure profile unchanged and verify if the data requires a different shape. 

We convolve the projected pressure profile with a Gaussian Beam of FWHM $2.4'$ relevant for CIB-deprojected Compton-$y$ maps \citepalias{Madhavacheril2020} to construct $y_{1h}$.

\begin{equation}\label{eq:y1h}
    y_{1h}(x_\perp) = \frac{\sigma_T}{m_e c^2}\int_{-\infty}^\infty P_e(\sqrt{x_\perp^2 + x_\parallel^2}) dx_\parallel \circledast Beam(x_\perp)
\end{equation}
\end{subequations}

We use the $y_{2h}$ from \cite{Vikram2017}, scaled for the average mass and redshift of our sample, and with variable amplitude, $A_{2h}$. 

Using the Affine-Invariant Ensemble Sampler algorithm implemented in \texttt{emcee} \citep{Foreman-Mackey2013}, we estimate the posterior probability distribution functions of $P_o$, $A_{2h}$, and $y_{zp}$. We set uniform priors of $-30\leqslant\rm P_o\leqslant30$, and $-10\leqslant\rm A_{2h}\leqslant 10$. Although negative values of these parameters are unphysical, allowing them to be negative keeps our analysis unbiased and lets us accurately constrain the upper limits in case\,of a non-detection. We assume the likelihood (e$^{-\chi^2/2}$; Equation\,\ref{eq:chisq}) to be Gaussian.

\begin{subequations}
    \begin{equation}\label{eq:chisq}
        \rm  \chi^2 =  [\overrightarrow{\rm Data} - \overrightarrow{\rm Model}(\mathbf{p})]^T   \mathbf{C_{\rm \bf stacked}}^{-1}[\overrightarrow{\rm Data} - \overrightarrow{\rm Model}(\mathbf{p})] 
    \end{equation}
    \begin{equation}
        {\rm where}\; \overrightarrow{\rm Data} = y\rm_{stacked}\; in\; Equation\,\ref{eq:stack},
    \end{equation}
    \begin{equation}\label{eq:model}
        \overrightarrow{\rm Model} = (y_{1h}+ A_{2h}y_{2h}+y_{zp}),
    \end{equation}
    \begin{equation}
        {\rm and}\; \mathbf{p} = [P_o, A_{2h}, y_{zp}]
    \end{equation}
\end{subequations}

We run multiple \texttt{emcee} ensembles starting from 5000 chains and keep on adding independent sets of chains until 1) the number of chains is larger than 50 times the integrated auto-correlation time, $\tau$, of each parameter, and 2) the Gelman-Rubin convergence parameter reaches values within 0.9--1.1 \citep{Gelman1992}. We remove the first 2$\tau_{max}$ steps from the sampler to get rid of the burn-in phase. We consider the most likely values of the free parameters, i.e., values corresponding to the minimum $\chi^2$, and the covariance matrices of their posterior probability distributions, \textbf{C$_{\rm \bf emcee}$}, for the following calculations. 

%To test if a flatter pressure profile describes the data better, we repeat the above analysis with the shape of ``1-halo" term replaced by the best-fit results of \citetalias{Amodeo2021}: $\alpha = 0.8, \beta = 2.6$ in Equation\,\ref{eq:pgnfw}. For convenience, we will refer to this model as ``flatter GNFW" and the model of \citetalias{Arnaud2010} as ``fiducial GNFW" in the following sections. 

%So far we have assumed that the shape of the pressure profile remains same across mass. 
%To test this, we replace GNFW profiles with the median and mean profiles predicted in simulations and refit the data. *to do*

\subsection{Thermal energy}\label{sec:Y}

For each mass bin, we calculate the thermal energy within a spherical volume of radius R$_{200}$ (Equation\,\ref{eq:Ysph}) from the best-fit pressure profiles in \S\ref{sec:P}. 
%a directly measurable model-independent quantity ${\rm Y^{cyl}} = \int y_{1h} \rm d\Omega$. $\rm Y^{cyl}$ converges to $\rm Y^{sph}$ if the ``1-halo" term becomes zero outside the radius of integration. 
\begin{subequations}
    \begin{equation}\label{eq:Ysph}
    {\rm \tilde Y^{sph}_{R200} = K} \int_{0}^{\rm R_{200}}{\rm P_{th}}(r)\,4\pi r^2 dr /(500\,\rm Mpc)^2 
\end{equation}
\begin{equation}\label{eq:K}
    {\rm where\;K} = (\sigma_{\rm T}/m_e c^2)E(z)^{-2/3}
\end{equation}
\begin{equation}\label{eq:Pth}
    {\rm and\;P_{th}}(r) = \Big(\frac{3\mu_e}{2\mu}\Big){\rm P_e}(r) 
\end{equation}
for a monoatomic ideal gas.
\end{subequations}
$\rm \tilde Y^{sph}_{R200}$ accounts for different redshifts of galaxies and normalizes thermal energy to $z=0$ through $E(z)$, and to a fixed angular diameter distance of 500\,Mpc. The choice of 500\,Mpc is inspired by literature but otherwise does not bear any physical significance.  

To test if our sample is consistent with self-similarity, we employ a hierarchical Bayesian model \citep[\href{https://github.com/jmeyers314/linmix}{linmix};][]{Kelly2007} for fitting a straight line to the relation between ln($\rm M_{200}$) and ln($\rm \tilde Y^{sph}_{R200}$), expecting that $\rm M_{200}$ and $\rm \tilde Y^{sph}_{R200}$ follows a power-law relation (Equation\,\ref{eq:YM}). %$\rm \tilde Y^{sph}_{R200} = Y_{200,o}(M_{200}/M_{200,o})^\Gamma$. 
\begin{equation}\label{eq:YM}
\begin{split}
        \rm \tilde Y^{sph}_{R200} = \tilde Y_{0}(M_{200}/M_{0})^\Gamma
\\
\implies {\rm ln(\tilde Y^{sph}_{R200})} = a_0 + a_1 \rm ln(M_{200})
\\
{\rm where}\; a_0 = {\rm ln(\tilde Y_{0})-\Gamma ln(M_{0})};\;\;{\rm and}\; a_1 = \Gamma
\end{split}
\end{equation}
$\rm M_{200}$ and $\rm \tilde Y^{sph}_{R200}$ are assumed to be drawn from a 2-d log-normal distribution $\mathcal{N}_2(\mu,\Sigma)$. Here, the mean $\mu = (\xi,\eta)$ represent the unobserved \textit{true} values of ln($\rm M_{200}$) and ln($\rm \tilde Y^{sph}_{R200}$), and the covariance matrix $\Sigma$ contains the measured errors of ln($\rm M_{200}$) and ln($\rm \tilde Y^{sph}_{R200}$). $\xi$ and $\eta$ are connected through ${\rm P(\eta|\xi)} = \mathcal{N}(a_0 + a_1 \xi, \sigma^2)$, where the regression parameters $a_0, a_1$, and $\rm \sigma^2$ denote the intercept, slope, and Gaussian intrinsic scatter of $\eta$ around the regression line, respectively. The priors on $a_0, a_1$, and $\rm \sigma^2$ are uniform. 

${\rm \tilde Y^{sph}_{R_\Delta} \propto \int^{R_\Delta}_0} P(r) r^2 dr$ (see equation\,\ref{eq:Ysph}) $\equiv \langle P \rangle R_\Delta^3$, where $P \propto n T$ and $R_\Delta^3 \propto M_\Delta$. In a self-similar relation, $\langle n\rangle \propto \Delta \rho_c$, and $\rm \langle T\rangle \propto (M_\Delta/R_\Delta) \propto M_\Delta^{2/3}$. Thus, $\rm \tilde Y^{sph}_{R_\Delta} \propto M_\Delta^{2/3} M_\Delta \propto M_\Delta^{5/3}$, implying the value of $\Gamma$ should be 5/3 if self-similarity is satisfied.

%Because different observational methods \citep[e.g., abundance matching;][]{Behroozi2010} might result in conflicting SHMR, the results of regression might be affected by 

%Because different observational methods \citep[e.g., abundance matching;][]{Behroozi2010} might result in conflicting SHMR, the results of regression might be affected by the SHMR considered in our analysis, which is based on weak lensing measurements and is different for quiescent and starfoming galaxies (see \S\ref{sec:galsample}). The SHMR is also different across theoretical simulations. To ease the comparison of our result with other observational studies and simulations, we repeat the regression analysis by replacing the virial mass with stellar mass, i.e., expecting that $\rm M_*$ and $\rm \tilde Y^{sph}_{R200}$ follows a power-law relation: $\rm \tilde Y^{sph}_{R200} = Y_{*,o}\times (M_*/M_{*,o})^\Gamma$. %Because the peak of SHMR generally occurs around the virial mass of 10$^{12}$\msun in the redshift range of our galaxies, the relation between stellar and virial mass in our considered range of 10$^{12-14}$\msun is essentially similar to a power-law. This validates our assumption made in this second regression analysis. 

\begin{figure*}
    \centering
\includegraphics[width=0.995\linewidth]{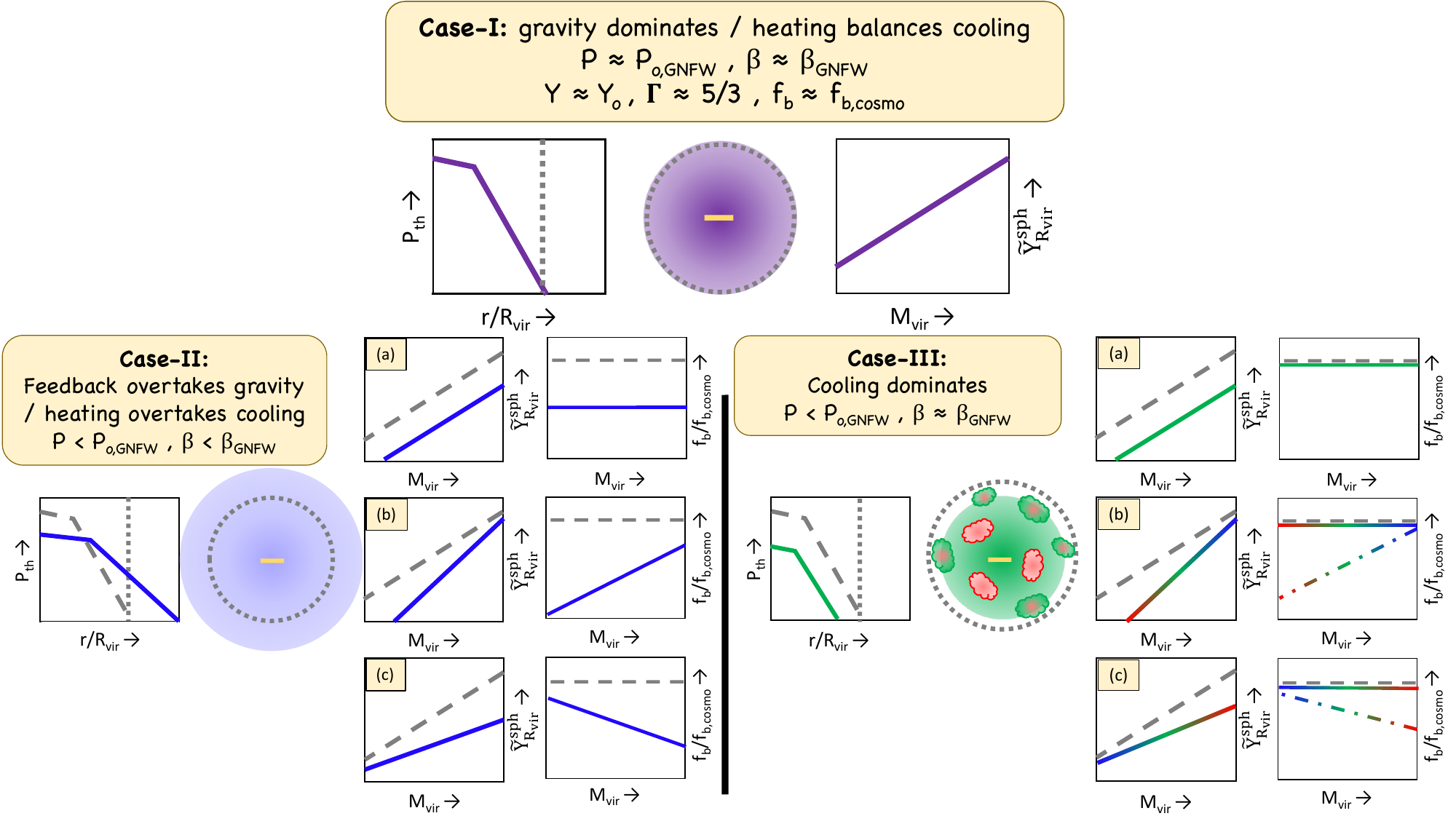}
    \caption{An illustration of possible physical scenarios. Diffuse ionized CGM probed by the tSZ effect is shown with colored circles, with purple-blue-green (case I-II-III) denoting gas temperature in descending order, and darker/lighter shades implying higher/lower gas density. Cooler and clumpy phases are not negligible in case\,III; those are shown with green/red clouds. The stellar disk is shown with a yellow bar for reference; the disk-to-halo is not drawn to scale. The virial radius is drawn with a gray dotted circle. In the thermal pressure profile ($\rm P_{th}$ vs r/$\rm R_{vir}$) plots, the gray dotted vertical lines mark the virial radius within which thermal energy ($\rm\tilde Y^{sph}_{R_{vir}}$) and baryon fraction ($f_b$) are calculated. The results of case\,I are shown with gray dashed lines in the plots of case\,II and case\,III. The multicolored lines in case\,III-b and III-c indicate variation in $\rm T/T_{200}$, with warmer/cooler gas shown in blue/red. $f_b \approx f_{b,cosmo}$ in case\,III (solid multicolored lines), but the contribution of the volume-filling CGM to $f_b$ would depend on the virial mass and cooling mechanisms; this is shown with dash-dotted lines. There is a degeneracy in the thermal energy-halo mass ($\rm\tilde Y^{sph}_{R_{vir}}$ vs $\rm M_{vir}$) relation between case\,II and III, which can be broken by considering the $\rm P_o$ and $\beta$ of the thermal pressure. See \S\ref{sec:scenario} for details.}
    \label{fig:cartoon}
\end{figure*}

\subsection{Baryon budget}\label{sec:masscal}

We calculate the mass of the hot CGM (Equation\,\ref{eq:mass}) assuming the volume-average temperature, $\rm \langle T_{CGM} \rangle$, to be the same as the virial temperature, $\rm T_{200}$. We estimate $\rm T_{200}$ from the virial mass and the virial radius (Equation\,\ref{eq:Tvir}) and calculate the density profile (Equation\,\ref{eq:nvir}) for the best-fitted pressure models considered in \S\ref{sec:P}.

\begin{subequations}
\begin{equation}\label{eq:mass}
   {\rm M_{CGM}} = \int_{0}^{\rm R_{200}} \mu_e m_p n_e(r) 4\pi r^2 dr E(z)^{-1}
\end{equation}
\begin{equation}\label{eq:nvir}
    {\rm where},\;\; n_e(r)  = {\rm P_e}(r) \rm k_B^{-1} T_{200}^{-1} 
\end{equation}
\begin{equation}\label{eq:Tvir}
   {\rm  and, T_{200}} = \frac{\mu m_p \rm G M_{200}}{\rm 2k_B R_{200}} 
\end{equation}
\end{subequations}
Here, $m_p$, $n_e$, and $\rm k_B$ are the rest mass of the proton, the number density of electrons, and Boltzmann's constant, respectively. Finally, we calculate the baryon fraction, $f_b = {\rm (M_\star + M_{CGM})/ M_{200}}$.% = \frac{\rm M_\star}{\rm M_{200}}(1+f_{gas,200})$.

Because the actual temperature of the hot CGM might differ from the virial temperature, we calculate the expected average temperature (Equation\,\ref{eq:TCGM}) if the galaxies were baryon sufficient. 

\begin{equation}\label{eq:TCGM}
   \frac{\rm \langle T_{CGM} \rangle}{\rm  T_{200}} = \frac{\rm M_{CGM}}{f_{b,cosmo}\rm M_{200} - M_\star}
\end{equation}

We calculate the mass of the gas traced by the tSZ effect and the mass of the dark matter (DM) halo within an arbitrary radius of $r = n\rm R_{200}$ and calculate $f_b(r)$. 
\begin{equation}
    f_b(r) = \frac{M_\star+M_{gas}(r)}{M_\star+M_{gas}(r)+M_{DM}(r)}
\end{equation}

At $r=R_{200}$, $f_b(r)$ converges to the standard definition of baryon fraction, $f_b$. 

\section{Results and discussion}\label{sec:result}

% for the fiducial GNFW pressure model and cyan squares and orange triangles for the flatter GNFW pressure model.  %$\rm \tilde Y^{sph}_{R200} \propto M_*^\Gamma$ (left) and  %$\frac{\rm \partial\tilde Y^{sph}_{R200}}{\rm \partial M_*}$ (left) 

\begin{figure*}
    \centering
    \includegraphics[width=0.995\linewidth]{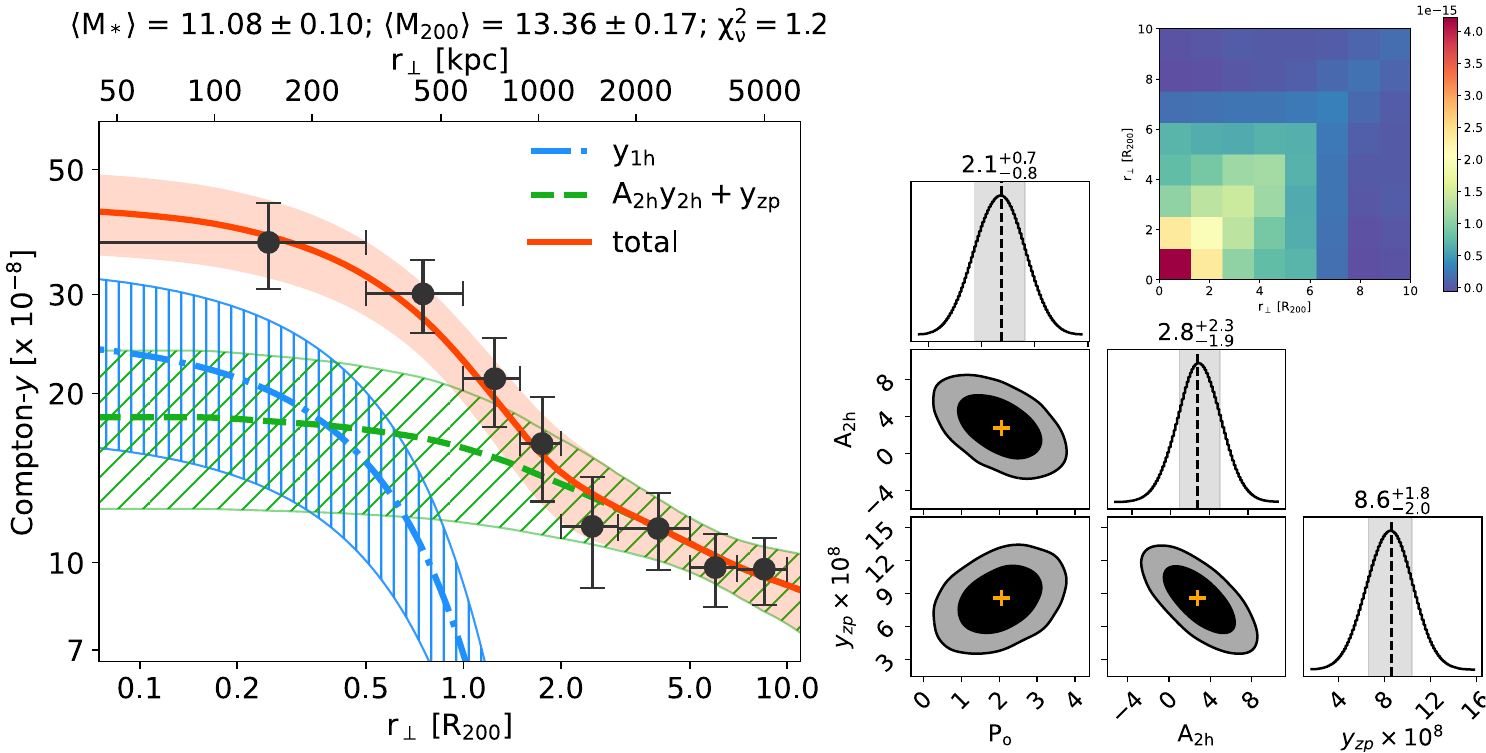}
    %\includegraphics[width=0.99\textwidth]{bestfit_13.2_to_13.7_Amodeo21.pdf}
    %%\vspace{-0.1 in}
    \caption{Results of stacking, aperture photometry, and pressure modeling for one of the mass bins of quiescent galaxies. %Top/bottom: \citetalias{Arnaud2010}/\citetalias{Amodeo2021} pressure profiles. 
    Results for all other mass bins are shown in Figure\,\ref{fig:all}. \textbf{Left:} Projected radial profile of Compton-$y$, $y_{stacked}$ (Equation\,\ref{eq:stack}). The top and bottom x-axes show the projected radius in the units of physical radius (kpc) and virial radius (\rvirialn). The width of each annular aperture (see Equation\,\ref{eq:ysub}) is shown with the error bars along the x-axis. The error bars along the y-axis are derived from the diagonal terms of the covariance matrix \textbf{C$_{\rm \bf stacked}$} (shown in the right panel). Note that these error bars are shown only for illustration; the full \textbf{C$_{\rm \bf stacked}$} is considered in the analyses. The dash-dotted blue curve and the dashed green curve are the best-fit ``1-halo" term, $y_{1h}$ and tSZ background (${\rm A_{2h}}y_{2h} + y_{zp}$), respectively, with the hatched area around them corresponding to the 1$\sigma$ uncertainty. The solid red curve and the shaded area around it are the best-fit models with 1$\sigma$ uncertainty. \textbf{Middle:} Posterior probability distributions of the amplitude of the thermal pressure profile: $\rm P_o$, normalization of the 2-halo term: $\rm A_{2h}$, and zero-point offset: $y_{zp}$, obtained by fitting the GNFW pressure model to our tSZ measurements. The vertical dashed lines and the shaded regions in the diagonal plots correspond to the most likely value and 68\% confidence interval of the marginalized distribution of the free parameters; the values are mentioned in the respective titles. The contour plots are drawn from \textbf{C$_{\rm \bf emcee}$}. The contours correspond to 68\% and 95\% confidence intervals; the most likely values are marked with `+'. \textbf{Right:} Covariance matrix of the $y_{stacked}$, \textbf{C$_{\rm \bf stacked}$}.}\label{fig:bestfit}
\end{figure*}
\subsection{Possible scenarios}\label{sec:scenario}
We discuss the general possible observable effects of gravity and major non-gravitational aspects in Figure\,\ref{fig:cartoon}. We consider six quantities: 
\\
1) the \textit{amplitude} ($\rm P_o$) and 
\\
2) the \textit{shape} ($\beta$) of the thermal pressure (see equation\,\ref{eq:pgnfw}),
\\
3) the \textit{normalization} ($\rm \tilde Y_o$) and 
\\
4) the \textit{slope} ($\Gamma$) of the thermal energy-halo mass relation (see equation\,\ref{eq:YM}), 
\\
5) the ratio of the volume-average temperature of the ionized CGM and the virial temperature ($\rm T/T_{200}$), 
\\
6) total baryon fraction with respect to the cosmological baryon fraction ($f_b/f_{b,cosmo}$), and the baryon content of the volume-filling ionized phase probed in the tSZ effect. 
\\
Below, we discuss three possible physical scenarios. Case\,I is the idealized scenario, and case\,II and case\,III are deviations from case\,I, primarily due to outward pressure and cooling, respectively. 
\begin{enumerate}%(cooling, thermal and kinetic mode feedback, non-thermal sources)
    \item \textbf{Case\,I:} If different non-gravitational aspects (e.g., heating and cooling) cancel each other's impact or have negligible impact in the first place, the ionized CGM properties are driven by gravity. In that case, $\rm P_o$ and $\beta$ would follow the standard GNFW profile. $\rm \tilde Y_o$ and $\Gamma$ would satisfy the self-similarity, $\rm T/T_{200}$ and $f_b/f_{b,cosmo}$ would be $\approx 1$ across mass. 
    \item \textbf{Case\,II:} In this scenario, the pressure from galactic feedback and/or non-thermal sources overcomes gravity. The kinetic mode feedback snowplows the CGM out to a large radius. Heating by thermal mode feedback and/or non-thermal sources outperforms cooling, and hence expands the ionized CGM. As a result, the thermal pressure deviates from the standard GNFW, with smaller $\rm P_o$ and $\beta$ than GNFW. $\rm \tilde Y_o$ and $f_b/f_{b,cosmo}$ would be smaller than those in case\,I within a fixed radius (e.g., \rvirialn), requiring a larger ``closure radius" to converge to case\,I. 
    \\ 
    Case\,II-a: If the relative strength of feedback to gravity remains the same across halo masses, $\Gamma$ would be similar to 5/3, and $f_b$ would be a similar fraction of $f_{b,cosmo}$ across mass. 
    \\
    Case\,II-b: If the feedback strength remains similar across masses, due to a weaker gravitational potential at smaller halo masses, more CGM would be spewed out, leading to $\Gamma> 5/3$. $f_b$ would be a decreasing fraction of $f_{b,cosmo}$. 
    \\
    Case\,II-c: If the feedback becomes weaker at smaller halo masses, $\Gamma$ would be smaller than 5/3, and $f_b$/$f_{b,cosmo}$ would increase at smaller halo masses, approaching the case\,I.
    \item \textbf{Case\,III:} In the absence of strong feedback, the halo would retain most/all of its baryons within the virial radius, and the multiphase structure of the CGM would be predominantly impacted by the cooling. Thus, combining all phases of the CGM, $f_b/f_{b,cosmo}$ would be $\approx 1$ across halo mass. Whether the volume-filling ionized CGM probed in the tSZ effect would be the most massive phase would depend on the halo mass and cooling efficiency. The volume filling phase could homogeneously cool to a sub-virial temperature, such that $\rm P_o$ would decrease with unchanged $\beta$. The $\rm \tilde Y_o$ would decrease due to decreased $\rm P_o$. 
    \\
    Unlike case\,II, where the primary effect is in the baryon content, temperature is the main affected property in case\,III. 
    \\
    Case\,III-a: If the cooling efficiency remains similar across halo masses, $\Gamma$ would be similar to 5/3. $\rm T/T_{200}$, and the baryon content of the volume-filling ionized CGM would be similar across halo masses. 
    \\
    Case\,III-b: If the cooling rate is higher at smaller halo masses, a fraction of the volume-filling phase might cool to clumpy phases. There, $\rm T/T_{200}$, and baryon content of the volume-filling phase would decrease, leading to $\Gamma>5/3$. \\
    Case\,III-c: If the cooling becomes weaker at smaller halo masses, $\rm T/T_{200}$, and the contribution of the volume-filling phase to $f_b$ would increase, leading to a $\Gamma < 5/3$, and approaching the case\,I. 
    %cooling is tricky because in massive halos cooled gas is still hot on absolute scale, but in lower masses cooled gas will be clumpy and missed in volume-average measurement like tSZ. 
    %\item Case\,IV: While not applicable to our galaxy sample, the CGM of satellite galaxies can be ram-pressure stripped by the halo gas of the central galaxy. The observable effect would be qualitatively similar to case\,II.   
\end{enumerate}

In the following sections, we identify the scenario apt for our galaxy sample based on the thermal pressure profile and the trend of thermal energy with halo mass. 

\subsection{Thermal Pressure}\label{sec:P_result}

In Figure\,\ref{fig:bestfit}, we show the results of stacking and aperture photometry for one of the mass bins of quiescent galaxies. The results for other mass bins are shown in Figure\,\ref{fig:all}. The S/N of the stacked $y$-value in the CGM ($r_\perp\rm\leqslant R_{200}$) varies from 4.9$\sigma$ to 10.4$\sigma$ for quiescent galaxies and 7.6$\sigma$ to 18.5$\sigma$ for star-forming galaxies. Because the data beyond the CGM is expected to be dominated by the tSZ background, the S/N of that part of the data does not bear any physical significance related to the signal, i.e., the ``1-halo" term. The S/N of quiescent galaxies is generally smaller than star-forming galaxies because of the smaller sample size. The S/N of quiescent galaxies increases with mass due to increasing signal strength. For star-forming galaxies, the data even within the CGM is dominated by the tSZ background, so the S/N increases with decreasing mass because of increasing sample size. 

In the left panel of Figure\,\ref{fig:bestfit}, we show the projected radial profile of stacked Compton-$y$ with best-fitted GNFW pressure models, the middle panels show the posterior probability distributions of the parameters of pressure models, and the right panels show the covariance matrices of the stacked Compton-$y$ profile. 

The ``1-halo" term (blue dash-dotted curve) and the tSZ background (dashed green curve) contribute with comparable strength in the CGM ($r_\perp\rm\leqslant R_{200}$), and the tSZ background dominates beyond the virial radius. We obtain good fits for the standard \textit{shape} of the GNFW model (red curve); the data (black points) do not require a flatter/steeper GNFW. This is consistent with case\,I and case\,III. However, the best-fitted value of the \textit{amplitude}, $\rm P_o$ ($\approx 2-3$), is smaller than the standard GNFW ($8.41$) at all considered masses (see in the titles of diagonal corner plots in Figure\,\ref{fig:bestfit} and \ref{fig:all}). Thus case\,III is preferred over case\,I. %Because we stack the $y$-profiles out to sufficiently large radius (10$\rm R_{200}$), our ``non-detection" of any excess CGM beyond the virial radius is quite robust. ($\chi^2/dof\approx1$) 

The excess in the data (black points) compared to the tSZ background (green hatched region) at $r_\perp\rm<R_{200}$ indicates the direct detection of the tSZ effect in the CGM. The best-fit values of the parameters imply a significant contribution of the ``2-halo" term and the zero-point offset (middle panel). $\rm A_{2h}$ and $y_{zp}$ are anti-correlated as expected for a given tSZ background. 

\begin{figure}[h] 
\centering
\includegraphics[width=0.995\linewidth]{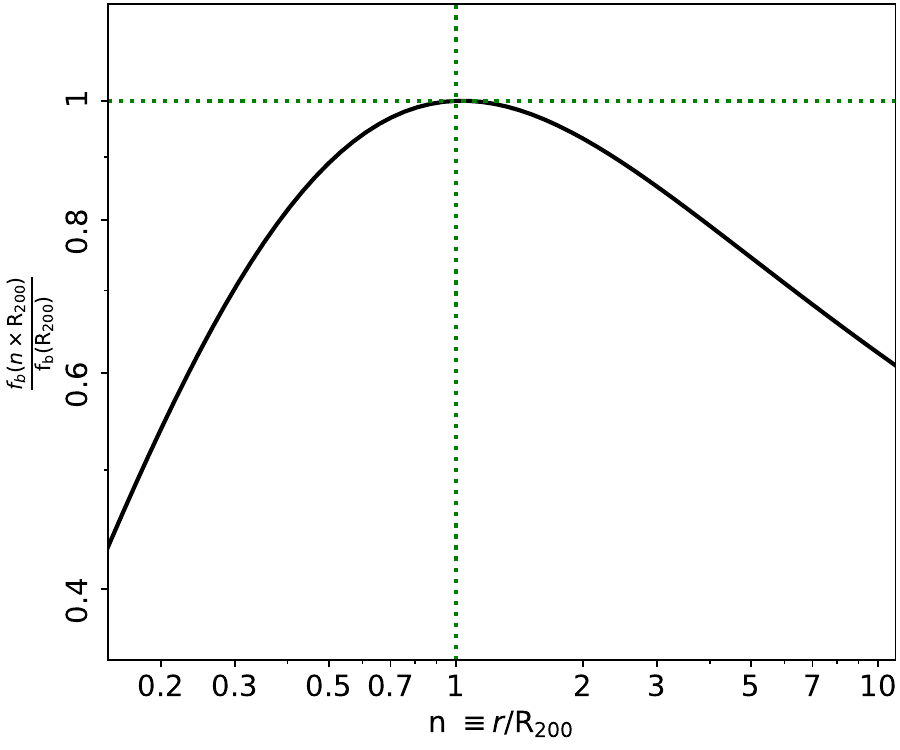}
%\vspace*{-2 in}
\caption{The ratio of the baryon fraction within an arbitrary radius to that within \rvirialn, for the standard GNFW model. The vertical and horizontal green dotted lines are drawn to guide the eye. See \S\ref{sec:fb_test} for details.}\label{fig:test_fb}
\end{figure} 

Beyond \rvirialn, we do not detect any excess in the data above the tSZ background (Figure\,\ref{fig:bestfit}, left panel). Because we stack the $y$-profiles out to sufficiently large radius (10$\rm R_{200}$), our ``non-detection" of any excess CGM beyond \rvirial is quite robust. It indicates that the ``1-halo" term is sufficient to explain the CGM emission. For the thermal pressure profile representing the ``1-halo" term, baryons beyond \rvirial do not contribute much to the baryon fraction (Figure\,\ref{fig:test_fb}). Therefore, either these halos are baryon-sufficient within \rvirialn, or, most of their \textit{galactic} baryons are thrown beyond 10\rvirial without leaving any observable trace between \rvirial and 10\rvirialn. Because the latter situation is physically unlikely, we infer that the galaxies in our sample where the tSZ effect is detected are baryon sufficient. 
%Because of the strong degeneracy among the three parameters of total thermal pressure, the flatter GNFW pressure profile with a weaker tSZ background fits the data equally well as the fiducial GNFW pressure profile with a stronger tSZ background. Thus, statistically, we cannot choose one pressure model over the other. In the following sections, we discuss how and why we prefer one of the models on physical grounds. 

\subsection{Self-similarity}\label{sec:self-sim-test}

\begin{figure}[h]  
\centering
\includegraphics[width=0.995\linewidth]{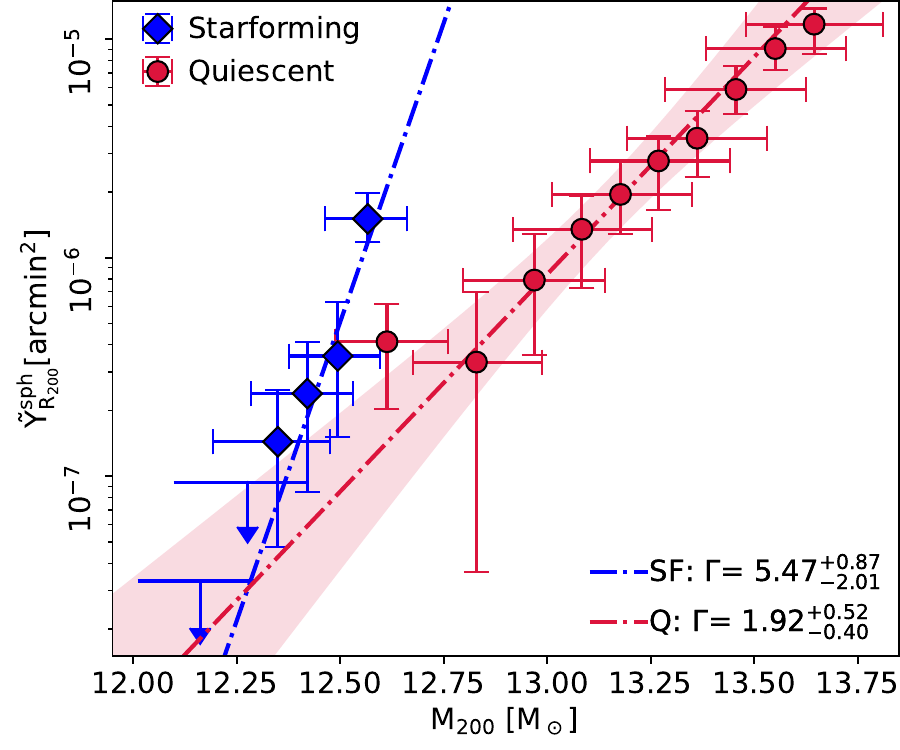}
\caption{$\rm \tilde Y^{sph}_{R200}$, a measure of the thermal energy within R$_{200}$, as a function of the halo mass, M$_{200}$. Observed values for star-forming and quiescent galaxies are shown with blue diamonds and red circles. The error bars along the x-axis correspond to the 1$\sigma$ uncertainties in the mean M$_{200}$ in the mass bin of consideration. The error bars in the y-axis show 1$\sigma$ uncertainties as well. The thermal energy is fitted with the model of $\rm \tilde Y^{sph}_{R200} \propto M_{200}^\Gamma$. The data for star-forming galaxies could not be constrained with this model due to a small number of mass bins. The dotted blue line shows an empirical estimate of $\Gamma$ calculated from the gradient, $\frac{\rm \partial\tilde Y^{sph}_{R200}}{\rm \partial M_{200}}$. The dashed red curve is the best-fit regression line for quiescent galaxies. The shaded area denotes the total (intrinsic + systematic) scatter in the regression. See \S\ref{sec:self-sim-test} for details. }\label{fig:YvsM}
\end{figure}

In Figure\,\ref{fig:YvsM}, we show the thermal energy of the CGM as a function of the virial mass. We detect the tSZ effect in the CGM down to the virial mass of 10$^{12.3}$\msunn. We show the best-fit values of $a_1$, i.e., the \textit{true} values of $\Gamma$, the \textit{slope} of the thermal energy-virial mass relation (see \S\ref{sec:Y}) for the two galaxy populations in the legend. 

In the CGM of quiescent galaxies, the value of $\Gamma$ is consistent with self-similarity within $1\sigma$. It is consistent with case\,I, II-a, and III-a. However, as discussed in the previous section, the \textit{amplitude} of the best-fitted thermal pressure is inconsistent with case\,I, and the  \textit{shape} of the best-fitted thermal pressure is inconsistent with case\,II. This leads us to choose case\,III-a as the most likely scenario.  

%The average thermal energy in the CGM for the flatter GNFW pressure is $3.0\pm0.6$ and $4.9\pm2.1$ times higher than that from the fiducial GNFW pressure for quiescent and star-forming galaxies, respectively. The relative dependence of thermal energy on mass (i.e., the value of $\Gamma$) is similar in both pressure models. %On the physical ground, we prefer the fiducial model of \citetalias{Arnaud2010} profile as a conservative estimate. 

%The dependence of thermal energy on stellar mass is similar for quiescent and star-forming galaxies within $1\sigma$. Thus, any difference in the dependence of thermal energy on virial mass would arise from the difference in SHMR of quiescent vs star-forming galaxies. 

Free-free cooling, which is usually expected to be the main cooling mechanism in the halo mass range of quiescent galaxies, is stronger at lower temperatures (and hence lower mass halos). The thermal energy-halo mass relation in that scenario would resemble case\,III-b. However, the observed relation follows case\,III-a. It indicates that there is subsidiary heating by thermal mode AGN feedback and/or non-thermal sources that would make up for more efficient cooling at lower halo masses. 

%Iimplies that feedback, if present, is not dominant over gravity, or the relative strength of feedback to gravity remains the same across the whole mass range of consideration. %This contradicts the strong AGN feedback implemented in simulations, throwing the gas outside the halo and creating cavities in the CGM. %It also contradicts the flatter pressure profiles extending beyond the CGM claimed in some previous observational studies \citep[e.g.,][]{Amodeo2021}. 

The thermal energy in the CGM of star-forming galaxies reveals a stronger dependence on virial mass than that expected in self-similarity. It is consistent with both case\,II-b and case\,III-b. Because the data do not require the thermal pressure \textit{shape} prescribed in case\,II-b (we verify it further in Appendix\,\ref{sec:flat}), we favor case\,III-b as the most likely scenario. It is consistent with the picture that metal line cooling, usually the main cooling mechanism in the halo mass range of star-forming galaxies, becomes stronger at lower temperature in lower mass halos. This also indicates an increasing requirement for non-thermal pressure support in lower mass star-forming halos to sustain them at hydrostatic equilibrium. 

The difference between quiescent and star-forming galaxies in the context of self-similarity also suggests that the broken power-law dependence of thermal energy on virial mass found in the galaxy sample of \citetalias{Das2023a} could be a manifestation of studying quiescent and star-forming galaxies together, with the more massive (and likely quiescent) galaxies following self-similarity but the lower mass end (likely star-forming) deviating from the self-similar relation. 

\subsection{CGM temperature}\label{sec:fb_test}

We cannot empirically determine the temperature profile of the volume-filling ionized CGM from the tSZ effect ($\propto\int n_e T_e dl$) alone; it is necessary to combine with the kinetic SZ (kSZ) effect, which is a measure of the density profile ($\propto\int n_e dl$). However, estimating the kSZ effect of this galaxy sample is not feasible with the current CMB data, as the kSZ effect is $\approx$an order-of-magnitude weaker than the tSZ effect. Despite this limitation, we can still infer the volume-average temperature of the ionized CGM, $\rm \langle T_{CGM}\rangle$, based on physical reasoning, thereby enabling more informed strategies about future tSZ + kSZ experiments. This exercise offers an estimate of \textit{``temperature floor"} for the ionized CGM, i.e., the minimum $\rm \langle T_{CGM}\rangle$ permitted by the tSZ data. Below this threshold, the implied CGM mass would render baryon-overabundant halos ($f_b>f_{b,cosmo}$), which is unphysical. Similarly, positing $\rm \langle T_{CGM}\rangle$ to not exceed the virial temperature (otherwise the entire halo will fall out of thermal and hydrostatic equilibrium - an unlikely condition for our galaxies, which are not individually identified as hosts of active nuclei\footnote{This scenario is different from radio-loud or generally active galaxies, where the effect of strong feedback is expected apriori and finding flatter density and/or pressure profile in SZ is reassuring rather than surprising.}), one can obtain the minimum CGM mass allowed by the tSZ data, thereby indirectly constraining the strength of ejective galactic feedback. Below, we discuss $\rm \langle T_{CGM}\rangle$ under the condition of a baryon-sufficient system in three distinct mass categories. %Because the best-fitted thermal pressure profile indicates baryon sufficiency for the galaxies with detected tSZ effect

\subsubsection{Higher mass}

Massive ($\rm M_{200} \geqslant 10^{12.8} M_\odot$) galaxies are baryon sufficient with their volume filling phase of the CGM at an average sub-virial temperature of 0.27$\pm$0.04$\rm \times T_{200}$ (Figure\,\ref{fig:fb}). Because of the relatively short cooling timescale in these halos, a cooled CGM is not unlikely in the absence of strong feedback, as discussed in case\,III in \S\ref{sec:scenario}. 

Because these galaxies are quiescent, to suppress the star formation, there has to be weak thermal mode AGN feedback (and/or non-thermal pressure) that prevents the inner CGM from runaway cooling, followed by infall to the stellar disk. It is in line with the discussion in the previous section on the role of subsidiary heating to maintain self-similarity. Given the large virial temperature of these galaxies, the CGM cooled at sub-virial temperature would still be at $\geqslant 10^6$\,K and fully ionized.

\subsubsection{Medium mass}

\begin{figure} 
\centering
\includegraphics[width=0.995\linewidth]{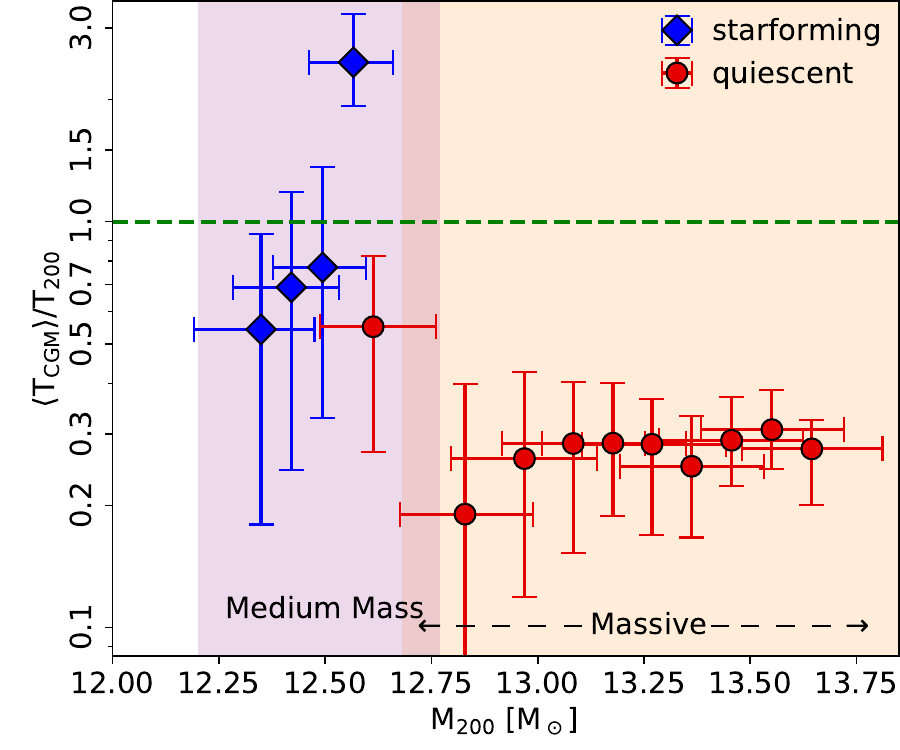}
%\vspace*{-2 in}
\caption{Volume-average temperature as a function of virial mass for a baryon-sufficient system. The horizontal dashed line denotes the virial temperature.}\label{fig:fb}%Baryon fraction (left) and v, the cosmological baryon fraction \citep[left]{Planck2020} and 
\end{figure}

Medium-mass galaxies ($\rm M_{200} \approx 10^{12.3-12.8} M_\odot$) are baryon sufficient with the volume filling phase of the CGM at their respective virial temperatures ($1.0_{-0.2}^{+0.3}\rm \times T_{200}$) of $\approx 10^{6.0-6.4}$\,K (Figure\,\ref{fig:fb}). The cooling timescale in these halos is generally large. Quiescent galaxies would be consistent with this picture. In star-forming galaxies, the CGM within a critical galactocentric radius could cool due to thermal instability and accumulate at the stellar disk \citep[e.g., see][]{Maller2004}, and get further heated by stellar feedback in the inner CGM to maintain the volume-average temperature at the virial value. As discussed in case\,III in \S\ref{sec:scenario}, it also rules out the possibility of any dominant galactic feedback. % 

%We prefer the second scenario on the physical ground.

%\subsubsection{Low mass galaxies ($\rm M_{200} < 10^{12.3} M_\odot$)}

\subsubsection{Lower mass}

We do not detect the ``1-halo" term in lower mass galaxies ($\rm M_{200} < 10^{12.3} M_\odot$; Figure\,\ref{fig:YvsM}). The non-detection has several implications for the volume-filling virial phase ($\approx 10^{5.9-6.1}$\,K): 
\\
1) Resolvability: Because the angular sizes of these halos are smaller than the ACT beam by a factor of $\gtrsim$3, the ``1-halo" term is inseparable from the tSZ background because of stronger beam dilution, and/or,
\\
2) Strong feedback: Case\,II-a is true; the thermal pressure profile is flatter than standard GNFW to the extent of being indistinguishable and hence inseparable from the ``2-halo" term in the fit. It would also explain systematically higher ``2-halo" term in these halos than those found in more massive halos (see Figure \ref{fig:all}), or, 
\\
3) Efficient cooling: The contribution of the volume-filling phase is negligible compared to the cooler and clumpy phases usually probed in UV absorption lines. %It would be consistent with the picture of efficient metal-line-dominated cooling at the virial temperature of these galaxies, and no strong thermal mode feedback that heats and ionizes the extended CGM. 

However, distinguishing among these three possibilities is currently impossible; more sensitive and higher angular resolution mm data are required. %, and then the cooling is suppressed due to overionization.  

%we compare thermal energy and actual baryon fraction (no temperature weight)

\begin{figure*}   
\centering
\includegraphics[width=0.995\linewidth]{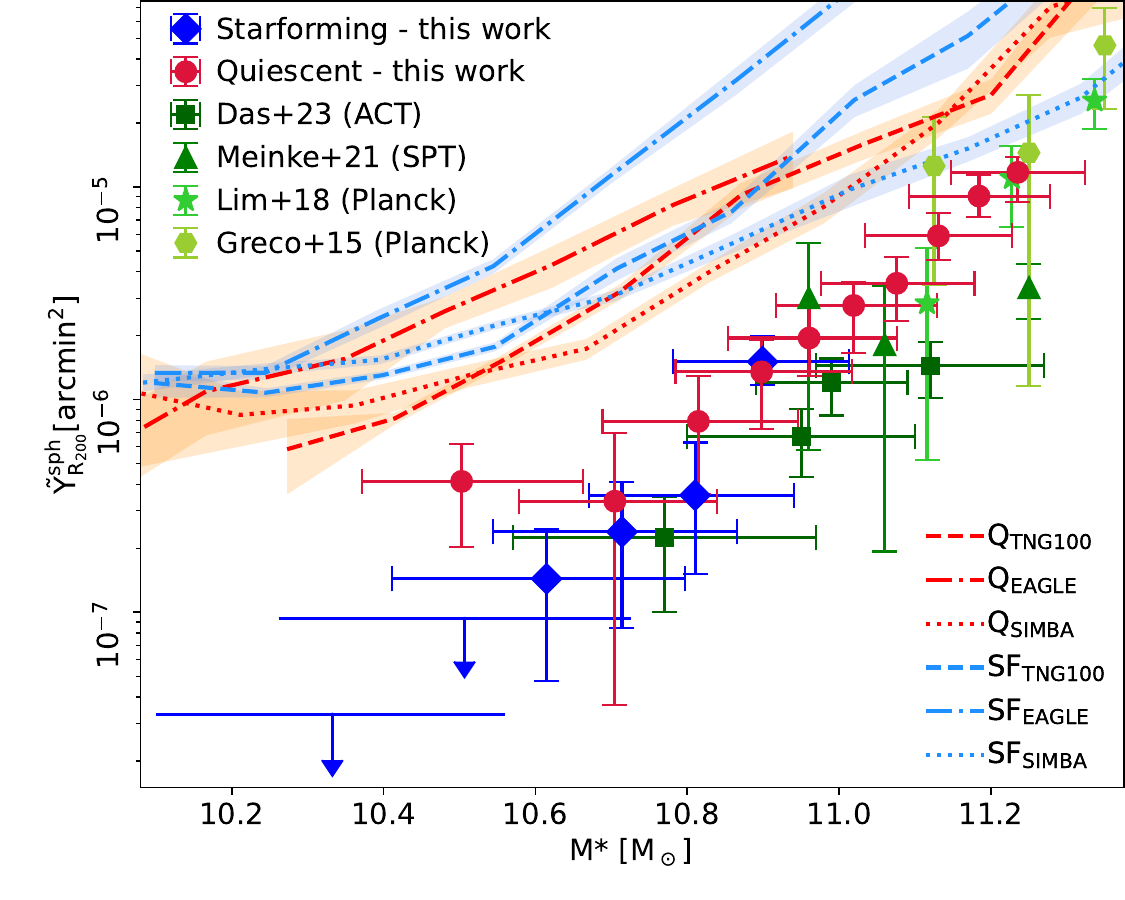}  
\caption{$\rm \tilde Y^{sph}_{R200}$, a measure of the thermal energy within R$_{200}$, as a function of the stellar mass, M$_\star$. The details about observed values are the same as Figure\,\ref{fig:YvsM}. Results from previous studies based on \planckn, ACT, and SPT data are shown with different green symbols. Our result is also compared against the mean thermal energy from simulations, with TNG100, EAGLE, and SIMBA shown with dashed, dash-dotted, and dotted curves, respectively. The shaded regions denote the uncertainty in the mean. See \S\ref{sec:compsim} for details.} \label{fig:YvsMstar}
\end{figure*} 

%In the top left panel of Figure\,\ref{fig:fb}, we show the baryon fraction in the CGM assuming a virialized system. 
%In the bottom panel of Figure\,\ref{fig:fb}, we show the ratio of $f_b(r)$ to $f_b$ as a function of the enclosing radius $r$. In the following, we discuss three categories of virial mass in the context of baryon sufficiency within \rvirialn. %: 1) quiescent galaxies, 2) star-forming galaxies with $\rm M_{200} \geqslant 10^{12.3}$\msunn, and 3) star-forming galaxies with $\rm M_{200} < 10^{12.3}$\msunn. 

%In Figure\,\ref{fig:fb}, we show the average temperature of the ionized CGM under the condition of a baryon-sufficient system. 

\subsection{Comparison with previous studies and simulations}\label{sec:compsim}

In Figure\,\ref{fig:YvsMstar}, we show the thermal energy of the CGM as a function of the stellar mass. We detect the tSZ effect in the CGM down to the stellar mass of 10$^{10.6\pm0.2}$\msunn, extending the detections down from the stellar mass of 10$^{10.8\pm0.2}$\msun in \citetalias{Das2023a}. Qualitatively, the thermal energy decreases with decreasing mass as expected, following a similar trend for star-forming (blue diamonds) and quiescent (red circles) galaxies. However, the virial mass (and hence the virial volume, $\rm V_{200}$) of quiescent galaxies is larger than that of star-forming galaxies at the same stellar mass (see \S\ref{sec:galsample}). Thus, similar thermal energy, $\rm \langle P_{th}\rangle V_{200}$, of both galaxy populations at the same stellar mass implies that the CGM of quiescent galaxies has weaker volume-average thermal pressure, $\rm \langle P_{th}\rangle$, than star-forming galaxies. Therefore, under the assumption of hydrostatic equilibrium, quiescent galaxies require stronger non-thermal pressure support than star-forming galaxies at the same stellar mass. 

The SHMR is often different across theoretical simulations and different observational studies, and could differ from the SHMR considered in our analysis (see \S\ref{sec:galsample}). To ease the comparison of our results with literature and predictions from simulations, we compare our thermal energy as a function of stellar mass. 

\subsubsection{tSZ effect}

The results from previous studies based on \planckn, ACT, and SPT data are shown in green symbols of different shapes: hexagons \citep{Greco2015}, stars \citep{Lim2021}, triangles \citep{Meinke2021}, and squares \citep{Das2023a}. %If only virial mass is quoted \citep[e.g.,][]{Lim2021}, we convert them to stellar mass using the SHMR of quiescent galaxies considered in our analyses. If the thermal energy is quoted within $\rm R_{500}$ \citep[e.g.,][]{Greco2015,Das2023a}, we convert that to the thermal energy within \rvirial using the GNFW pressure profile considered in our analyses.  
It should be noted that comparing different observational studies is nontrivial because of their differences in sample selection, masking, process of $y$-value extraction, and thermal energy calculation. We exclude cluster members and AGNs from our galaxy sample, while some studies explicitly chose active central galaxies. We stack Compton-$y$ maps constructed through component separation, while some studies extract the Compton-$y$ value after channel-wise stacking. We mask regions of high Galactic dust extinction, tSZ-detected galaxy clusters, and radio-loud sources \citepalias[see each of their effects on stacking in][]{Das2023a}, which has not been considered in most previous studies. We cross-correlate galaxy catalogs with Compton-$y$ maps that make our stacked Compton-$y$ profile model-independent and give us more freedom in exploring various thermal pressure profiles. On the other hand, some studies employ matched filters of presumed pressure profiles. We forward model the signal (``1-halo" term) and background (``2-halo" term and zero-point offset) together, and integrate the thermal energy density corresponding to the signal within a spherical volume, while some studies quote direct integrals of stacked $y$-value within a cylindrical volume - making the separation of tSZ background challenging. Despite all these differences, previous results are broadly consistent with each other (modulo some scatter) and our work within 1$\sigma$ uncertainties, showing the synergy across different datasets and galaxy populations.  

\subsubsection{X-ray emission (stacking)}

We compare our results with the 0.5-2\,keV rest-frame luminosity of the CGM of $z<$0.1 stacked SDSS galaxies within $\rm R_{500}$ measured with eROSITA \citep{Zhang2025}. Quiescent galaxies are systematically X-ray brighter than star-forming galaxies at similar masses in their work, in apparent contradiction with our result. 

However, there are several fundamental differences between X-ray emission and the SZ effect. Accounting for the spatial and temporal variability of the multi-component soft X-ray background (SXB) dominating the total X-ray emission is extremely challenging in stacking analyses, as multiple galaxies are observed at different times and in different parts of the sky. Thus, contamination by residual SXB in a stacked signal is not impossible. Secondly, X-ray emission of the CGM is dominated by highly ionized metal lines, naturally resulting in a strong dependence on chemical composition. It also causes a non-monotonic measurement bias in temperature toward values with higher emissivity of tracer metals. Thirdly, X-ray emission is biased toward denser gas due to the density-squared dependence. Thus, it primarily probes the inner CGM. On the other hand, the tSZ effect, without any metallicity dependence and a linear dependence on the temperature and density, probes all the ionized baryons together, including lower-density gas extended out to larger radii. Thus, even for a genuine detection of CGM emission in X-rays, the unavoidable bias in density, temperature, and metallicity could be the reason behind the difference between our work and X-ray emission. 

Moreover, our target selection is different. The galaxy sample considered in X-ray emission is at $z<0.1$, unlike our sample extended out to $z=1.2$. We have corrected for the cosmology, but there could be an intrinsic difference in the CGM properties at different redshifts due to evolving feedback, cooling, and clustering history. The halo mass- and clustering-corrected redshift evolution of our galaxy sample would be explored in a follow-up paper (Das \textit{et al.} 2025b, in prep.). Also, we exclude known AGNs from our sample, but an unknown fraction of the X-ray galaxy population could host an AGN. Contamination in the CGM by the average nuclear emission (convolved by eROSITA PSF) is accounted for, but the AGN outflows could still enhance the measured X-ray luminosity of the CGM. A joint X-ray-SZ stacking analysis on the same galaxy sample should be ideal, but the flux limitation of X-ray and volume limitation of SZ currently restricts our ability for an effective experiment; this could be achieved with more sensitive X-ray and mm survey data in the future. Lastly, galaxy halos in tSZ are not resolved, limiting the constraint on the \textit{true} shape of the pressure profile. The pressure profile may be different for quiescent and star-forming galaxies, with the former having larger/smaller thermal pressure within/beyond $\rm R_{500}$ than the latter. It would enhance the X-ray luminosity of quiescent galaxies within $\rm R_{500}$ but provide similar thermal energy for both kinds of galaxies within \rvirialn. Testing this would require higher angular resolution mm data. %is beyond the scope of this paper.   %and the current mm and X-ray data. 
%Fourthly, assessing the ``2-halo" term is more complicated in X-ray due to the point source (AGN+XRB) contribution.  The clumpy CGM along the line of sight at a given impact parameter. different redshift.

\subsubsection{EUV absorption, X-ray absorption and emission} %, or, all that is not gravity is not feedback}

Previously, detection/non-detection of \ovi absorption line against background quasars has been reported in the CGM of star-forming/quiescent galaxies \citep[e.g.,][]{Tumlinson2011}. However, star-forming and quiescent galaxies in that sample were of low and high stellar mass, respectively. Thus, it was inconclusive whether the trend of \ovi detection was because of mass or star-forming activities. The virial mass of our galaxy sample spans $\approx$ two orders of magnitude, and the tSZ effect can probe any volume-filling ionized phase. Thus, we are sensitive to a broader range of temperature than \ovin, which traces 10$^{5.5}$\,K collisionally ionized gas (but with a peak ionization fraction of $f_{\rm OVI} \approx 0.2$ it is still dominated by lower and higher ionization states of oxygen, i.e., \ov and \oviin, at that temperature, with $f_{\rm OV} \approx f_{\rm OVII} \approx 0.4$). Learning from our tSZ observations, we infer the following about the \ovi observations.

Our sample is fitted well with the standard GNFW thermal pressure profile, without any evidence of excess gas beyond \rvirial out to 10\rvirialn. Thus, these galaxies are likely baryon sufficient within \rvirial (see \S\ref{sec:P_result}). In massive (super-L$^\star$) and medium-mass (L$^\star$) galaxies, the most massive phase of the CGM is volume-filling and at the sub-virial and virial temperatures of $\gtrapprox 10^6$\,K, respectively (see \S\ref{sec:fb_test}). This phase can be probed by \ovii and \oviii lines in X-ray (absorption and emission) as has been extensively studied in the CGM of Milky\,Way \citep[e.g.,][]{Nicastro2002,Henley2010,Gupta2012,Das2021a} and a few external galaxies \citep{Das2019b,Das2020a,Mathur2021,Nicastro2022,Mathur2023}. Therefore, galaxies of these masses are unlikely to show wide evidence of strong \ovi lines. In the least massive (sub-L$^\star$) galaxies in our sample, the CGM could cool to an \ovin-dominated clumpy phase and hence cannot be efficiently probed by the tSZ effect. Thus, assuming the galaxies of \cite{Tumlinson2011} and our sample are similar, the dichotomy in \ovi detection is driven by the mass, and consistent with theoretical predictions of \cite{Oppenheimer2016}. This also reveals that the apparent non-monotonicity in the baryon fraction (assuming virial temperature) found in the galaxies of \citetalias{Das2023a} was likely due to the non-monotonicity in the volume-average temperature of the ionized CGM. 

%caveats
%dust in the center suppressing y1h. brighter halo in the bkg enhancing y1h. satellites giving a "flatter" appearance. brighter halo in projection enhancing y2h. could 2h have a different shapes for starfoming and q galaxies? 
%we do not excludethe  innermost radial bin, unlike others. thus a conservative estimate of y.

\subsubsection{Simulations}

We compare the observed thermal energy with predictions from the three cosmological hydrodynamical simulations: TNG100 from IllustrisTNG \citep{marinacci.etal.2018,naiman.etal.2018,pillepich.etal.2018,springel.etal.2018,Nelson2019}, EAGLE \citep{crain.etal.2015, schaller.etal.2015, Schaye2015, Mcalpine.etal.2016}, and SIMBA \citep{dave.etal.2019}. The simulations are designed to explore the formation and evolution of galaxies within a $\Lambda$CDM cosmological framework, using consistent cosmological parameters and a simulated box of $\approx$100 Mpc per side. The three simulations incorporate key astrophysical processes relevant to galaxy formation and evolution, e.g., radiative cooling, star formation, stellar and AGN feedback, but they differ in the detailed implementation of these processes. It is important to note that these simulations are all calibrated to reproduce realistic galaxy populations at low redshifts ($z\sim0$). The properties of the hot CGM are pure predictions from these models, making them appropriate for meaningful comparison with observations.

The simulated galaxy samples are drawn from snapshots with redshifts comparable to the median observed redshifts. Specifically, the redshifts of simulated quiescent (star-forming) galaxies are 0.62 (0.73), 0.62 (0.74), and 0.63 (0.72) for TNG100, EAGLE, and SIMBA simulations, respectively.

We compare the mean thermal energy of the CGM of 661, 143, and 924 quiescent galaxies and 1245, 1068, and 2630 star-forming galaxies in TNG100, EAGLE, and SIMBA, respectively, with our result. The thermal energy of star-forming galaxies is overestimated in all simulations by more than a factor of 40. The discrepancy increases with decreasing mass. In quiescent galaxies, TNG100, EAGLE, and SIMBA overestimate the thermal energy by a factor of 9, 14, and 7, respectively. 

The stacking procedure estimates the average by construction and should be compared with the mean of the thermal energy distribution in simulations. Nonetheless, we consider the median from the simulations because the true thermal energy distribution might not be sampled by the small number of galaxies in the simulations, resulting in a skewed distribution. The median is systematically lower than the mean in all simulations for both quiescent and star-forming galaxies. However, the median thermal energy of star-forming galaxies is still a factor of 30 or more higher in simulations than in our observations. The median thermal energy of quiescent galaxies in TNG100 and EAGLE is overestimated by a factor of 6 and 9, respectively, while SIMBA is consistent with our observations within a factor of 3. 

Generally, EAGLE implements the weakest feedback, and SIMBA exhibits the strongest feedback. Thus, at first glance, it might seem like our galaxy sample prefers a stronger feedback scenario. However, the systematically lower thermal energy in observations does not necessarily indicate an effect of feedback (see \S\ref{sec:scenario}). Stronger feedback produces a flatter pressure profile through heating and snowplowing, with a non-negligible fraction of the CGM outside \rvirialn, which is not required by the current data. A cooler and/or less dense CGM in its volume-filling phase would also result in lower thermal energy, while retaining most of its baryons (across phases) within \rvirialn; this is consistent with the current data. Given the overall disagreement between observational data (as reported in our work and literature) and simulations, we abstain from commenting further to prevent any potential over-interpretation. 

Because star-forming galaxies are dustier than quiescent galaxies, the thermal energy of star-forming galaxies might have been underestimated in observations due to the residual effect of dust from their ISM/CGM. This could be the reason for a larger discrepancy between simulation and observation for star-forming galaxies than for quiescent galaxies. The angular resolution, sensitivity, and the dominant ``2-halo" term currently restrict our ability to assess this effect and correct for it in observations; this would be a part of a future endeavor with next-generation mm data.        

%Higher angular resolution $y$-maps will be able to resolve the halos and constrain the shape of the thermal pressure  
% Quiescent galaxies follow a self-similar relation (Figure\,\ref{fig:YvsM}), indicating that the relative strength of feedback to gravity remains similar across mass.

%Question to Nhut and others: The simulation with a larger number of samples is closer to the data. In addition to the underlying physics, could the number of samples affect the comparison? %%%--> Nhut: The size of the simulated samples may affect the results, in particular at the high-mass end, in which simulations with larger simulated boxes, e.g., SIMBA, produce more massive galaxies. However, based on the intrinsic scatters of the simulated Y_200-M* relations, I don't think the sample size could be an issue here. The simulated physical models may play a dominant role here. 

% for the fiducial GNFW pressure model and cyan squares and orange triangles for the flatter GNFW pressure model.  %$\rm \tilde Y^{sph}_{R200} \propto M_*^\Gamma$ (left) and  %$\frac{\rm \partial\tilde Y^{sph}_{R200}}{\rm \partial M_*}$ (left) or 

\section{Summary and future directions}\label{sec:end}
%Reworded. SM.
In \citetalias{Das2023a} we presented our results on the cross-correlation of 0.6 million $z\approx 0-0.3$ galaxies in the WISE$\times$SuperCosmos catalog with the Compton-$y$ map from ACT\,DR4. In this paper, we extend that study by leveraging the same $y$-map and cross-correlating it with 0.7 million quiescent and 1.8 million star-forming galaxies of $\rm M_{200}= 10^{12-14} M_\odot$ at $z=0.01-1.19$ in the WISE$\times$DESI catalog. We explore the dependence of the tSZ effect on the star-forming activities for the first time. We perform the same rigorous procedure of sample selection (i.e., no cluster member, IR-selected AGN, or radio-loud source) and $y$-map cleaning (from cosmic and Galactic dust, galaxy clusters, and radio sources) following \citetalias{Das2023a}. Additionally, we have devised and employed a novel stacking method that fully takes into account the uncertainties in redshift, mass, and star formation rate. This new method stacks Compton-$y$ values at a fixed fraction of the virial radius instead of the conventional stacking at fixed angular size or physical radius, and does not require any $y$-map reconstruction; thus, it saves computational time, data storage, bypassing the interpolation-related uncertainties in image reconstruction, and producing more accurate results and straightforward interpretation. We summarize our science results in the following.

\begin{enumerate}
    \item Detection significance of the stacked Compton-$y$ values in the CGM varies from 4.9$\sigma$ to 18.5$\sigma$ in different mass bins, before separating the signal (``1-halo" term) from the background (``2-halo" term and zero-point offset). 
    \item We fit the data with a standard GNFW thermal pressure profile. The model fits the data well, and a flatter/steeper profile is not required. It indicates the absence of any dominant feedback; it is the cooling, non-thermal sources, and subsidiary (thermal mode) feedback that shapes the CGM of these galaxies.  
    \item We can extract the signal down to the stellar mass of $10^{10.6\pm0.2}$\msunn, extending the detections down from $10^{10.8\pm0.2}$\msun in \citetalias{Das2023a}.
    \item The thermal energy in the CGM of quiescent galaxies follows the self-similar relation, but the star-forming galaxies do not, indicating a different impact of non-gravitational factors, e.g., feedback and cooling in quiescent vs star-forming galaxies. This also explains the broken power-law dependence of thermal energy on virial mass found in \citetalias{Das2023a} that studied quiescent and star-forming galaxies together due to a smaller sample size. 
    \item  If the volume-filling ionized CGM probed by the tSZ effect is the most massive phase, and on average is at sub-virial ($\approx 0.27\pm0.04 \times \rm T_{200}$) and virial temperatures of $\gtrsim 10^6$\,K in $\rm M_{200} >10^{12.8} M_\odot$ and $\rm M_{200} = 10^{12.3-12.8} M_\odot$ galaxies, respectively, they could be baryon sufficient.
\end{enumerate}

We have performed cross-correlations with photometric redshift catalogs to obtain a large galaxy sample. This allows us to sufficiently reduce the statistical uncertainty in the stacked Compton-$y$ value. However, this comes at the cost of large uncertainty in the redshift and other physical parameters derived from/dependent on it. Also, due to the intrinsic scatter in the CGM properties of similar galaxies, the systematic uncertainty dominates the total uncertainty in the stacked Compton-$y$ value of such a large galaxy sample. Galaxy surveys with \textcolor{black}{Rubin, Roman, and Euclid will extraordinarily increase the sample size of galaxies with photometric redshifts measured more precisely than the current standard.} % and of comparable quality to current spectroscopic redshifts
Upcoming/proposed telescopes, e.g., Simons Observatory, AtLAST, CMB-HD, CCAT, LiteBIRD, etc., will improve the quality of the SZ signal with increased sensitivity, footprint, angular resolution, and frequency coverage. This will allow us to achieve statistical uncertainty similar to current cross-correlation studies with smaller galaxy samples and thus obtain better control of the systematics. It will also enable us to resolve the CGM better and constrain the shape of the thermal pressure profile, and measure the kSZ effect alongside to constrain the density and pressure profiles independently, leading to a better understanding of the effect of galactic feedback and cooling on the ionized CGM.  

\section*{Data availability}
Our analyses use a publicly available galaxy catalog, $y$ maps, and simulated datasets. The stacked $y$ profiles, results of pressure modeling, and derived quantities from those are available \textcolor{black}{on our \href{https://github.com/sanskritiastro/thermal_SZ_II}{GitHub repository}}. %from the corresponding author upon reasonable request.  

\section*{acknowledgments}
\textcolor{black}{We thank the anonymous referee for encouraging comments and constructive suggestions.} S.D. acknowledges support provided by NASA through Hubble Fellowship grant HST-HF2-51551.001-A awarded by the Space Telescope Science Institute, which is operated by the Association of Universities for Research in Astronomy, Inc., for NASA, under the contract NAS 5-26555. S.D. acknowledges support from the KIPAC Fellowship of Kavli Institute for Particle Astrophysics and Cosmology, Stanford University, as well. N.T. acknowledges the support by NASA under award number 80GSFC24M0006. Y.C. acknowledges the support of the National Science and Technology Council of Taiwan through grant NSTC 111-2112-M-001-090-MY3 and Academia Sinica through the Career Development Award AS-CDA-113-M01. S.M. is grateful for the NASA ADAP grant 80NSSC22K1121. This work was initiated/performed/performed in part at the Aspen Center for Physics, which is supported by a grant from the Simons Foundation (1161654, Troyer). This research has made use of NASA's Astrophysics Data System Bibliographic Services. 

S.D. dedicates this paper to her father, Ashoke Kumar Das, who passed away during the preparation of this paper, and whose resilience and incredible sacrifices have been (and will always be) instrumental to all of her academic endeavors.

%This publication makes use of data products from the Wide-field Infrared Survey Explorer (WISE), which is a joint project of the University of California, Los Angeles, and the Jet Propulsion Laboratory/California Institute of Technology, funded by the National Aeronautics and Space Administration.

\facilities{\textit{DECaLS}, \textit{WISE}, \planckn, \actn, \jvla}
\software{\texttt{AstroPy} \citep{Astropy2022}, \texttt{corner} \citep{Foreman-Mackey2016}, \texttt{emcee} \citep{Foreman-Mackey2013},  \texttt{Halotools} \citep{Hearin2016}, \texttt{Jupyter} \citep{jupyter2016}, \texttt{Matplotlib} \citep{Hunter2007}, \texttt{NumPy} \citep{numpy2020}, \texttt{pandas} \citep{pandas2020},  \texttt{Pixell} \citep{Naess2021}, \texttt{Python math} \citep{mathpy2020},  \texttt{SciPy} \citep{scipy2022}}
%\texttt{scikit-image} \citep{scikitimage2014},

%A handy "cheat sheet" that provides the necessary \latex\ to produce 17  different types of tables is available at \url{http://journals.aas.org/authors/aastex/aasguide.html#table_cheat_sheet}.

\appendix 
\counterwithin{figure}{section}

\section{Stacking and aperture photometry}\label{sec:stack-technical}

%We develop a novel probabilistic approach in stacking the $y$-stamps by fully accounting for the uncertainties in redshift, SFR, and halo mass.  

Below, we discuss the procedure of stacking and aperture photometry step-by-step.

\begin{subequations}
\textbf{Step 1.} Using $\rm SFR_{MS}$ from Equation\,\ref{eq:sfms}, we calculate the probability that a galaxy is star-forming, $p_{\rm SF}$, as 
\begin{equation}
    p_{\rm SF} = \frac{1}{\sqrt{2\pi}\sigma_x}\int_{-0.6}^{\infty} e^{\frac{(x-\mu_x)^2}{2\sigma_x^2}} dx 
\end{equation}
\noindent Here, $\mu_x$ and $\sigma_x^2$ are the mean and variance in $`x$', $\rm log_{10}(SFR/SFR_{MS})$. The uncertainties in stellar mass and SFR quoted in the galaxy catalog \citepalias{Zou2019} are propagated to calculate the variance.

\textbf{Step 2.} Employing the results from step 1, we calculate the normalized probabilities of a galaxy to be star-forming or quiescent, $w_{\rm SF}$ or $w_{\rm Q}$, using Equation\,\ref{eq:psf}.
\begin{equation}\label{eq:psf}
    w_{\rm SF} = \frac{p_{\rm SF}}{\sum p_{\rm SF}} ;\; w_{\rm Q} = \frac{p_{\rm Q}}{\sum p_{\rm Q}} ;\;  p_{\rm Q} = 1 - p_{\rm SF} 
\end{equation}

\textbf{Step 3.} We split our sample into 40 halo mass \textit{slices} of width $\rm \Delta M_{200}$ = 0.05\,dex, and 60 redshift \textit{slices} of width $\Delta z$ = 0.02\,dex. This creates a 2-D grid of 2400 \textit{cells}  covering the entire range of halo mass and redshift. We calculate the probability of a galactic property $`x$', halo mass or redshift, in the $k$-th \textit{slice} of width $\Delta x$ as
\begin{equation}
   p(x_k) =  \frac{1}{\sqrt{2\pi}\sigma_x}\int_{x_k-0.5\Delta x}^{x_k+0.5\Delta x} e^{\frac{(x- \mu_x)^2}{2\sigma_x^2}} dx
\end{equation}
Here, $\mu_x$ and $\sigma_x^2$ are the mean and variance in $`x$'. For redshift, the measured values and uncertainties are directly adapted from \citetalias{Zou2019}. The SHMR is used to derive halo mass and its uncertainty from stellar mass and its uncertainty quoted in \citetalias{Zou2019} (see \S\ref{sec:galsample}).

\textbf{Step 4.} Using $p({\rm M}_{200,i})$ calculated in step 3, we calculate the normalized probability of a galaxy to have a halo mass in the $i$-th mass \textit{slice},  $w(\rm M_i)$, using Equation\,\ref{eq:pmass}.

\begin{equation}\label{eq:pmass}
   w({\rm M}_{200,i}) = \frac{p({\rm M}_{200,i})}{\sum_i p({\rm M}_{200,i})} 
\end{equation} 

Different SHMR for star-forming and quiescent galaxies results in different $p({\rm M}_{200,i})$, so we calculate two separated sets of $w({\rm M}_{200{\rm-SF},i})$ and $w({\rm M}_{200{\rm-Q},i})$. 

\textbf{Step 5.} Using $p(z_j)$ calculated in step 3, we calculate the normalized probability of a galaxy to be in the $j$-th redshift \textit{slice},  $w(z_j)$, using Equation\,\ref{eq:pz}. 

\begin{equation}\label{eq:pz}
     w(z_j) = \frac{p(z_j)}{\sum_j p(z_j)} 
\end{equation}

%We create a mock galaxy catalog of the same size as our sample in each patch, BN, and D56, with the sky coordinates drawn from a 2D random distribution that spans the range of sky coordinates of our sample. and null $y$-stamps ($y_{null}$) both %($y$-stamp extracted around these mock galaxy positions)

\textbf{Step 6.} We extract the cutout of the $y$-map around each galaxy, which we call $y$-stamp, using the \texttt{thumbnail} subroutine of the \texttt{reprojection} module of \texttt{pixell} \citep{Naess2021}\footnote{Each $y$-stamp is projected onto a local tangent plane to remove the effect of shape and size distortions}. To ensure the tSZ background is well characterized in a signal-free region, we extract $y$-stamps out to a large projected distance, 10$\times$ \rvirialn, in respective halo mass \textit{slices}. Thus, for the $cell_{ij}$, i.e., the intersection of $i$-th halo mass \textit{slice} and $j$-th redshift \textit{slice}, the angular size of all $y$-stamps would be ${\rm 10R}_{200,i}/{\rm D}_{{\rm A},j}$\footnote{${\rm D}_{{\rm A},j}$ is the angular diameter distance at $z_j$}. 

\textbf{Step 7: radial profile$\rightarrow$stack instead of stack$\rightarrow$radial profile.} Conventionally, where the uncertainties in galactic properties are not considered while stacking, $y$-stamps of fixed angular size for each galaxy are directly stacked ($\sum_i y_i(\theta)$), and then the cumulative radial profile in circular apertures is calculated from the stacked $y$-stamp. Because in our case, the size of $y$-stamp and hence the number of pixels are different across \textit{cells}, we cannot directly stack them. 

%This issue of $y$-stamps of different sizes happens for a different reason from that in \citetalias{Das2023a}. 
The $y$-stamps are of different sizes in \citetalias{Das2023a} as well, but for a different reason. There, we extracted the $y$-stamp of the $i$-th galaxy out to $\theta_i = {\rm 8R}_{200,i}/{\rm D}_{{\rm A},i}$. Because the value of $\theta_i$ differed across galaxies, this made the size of $y$-stamps different. Using 2-D interpolation, we reconstructed each $y$-stamp to the median angular size in each mass bin, and then stacked them ($\sum_i y_i(median\;\theta_i)$). On the other hand, in this paper, the size of $y$-stamp is the same across galaxies in each \textit{cell}, but it differs across \textit{cells} that we have to stack upon. %we did not consider the uncertainties in galactic properties, but

%In the end, the 2-D distribution of Compton-$y$ values is not relevant in this paper, therefore we 
%I'm trying to remove "not relevant" here.  and then stack those radial profiles
We extract the projected radial profiles of Compton-$y$ from each $y$-stamp in each \textit{cell}. It saves computational time and interpolation-related uncertainties in $y$-stamp reconstruction and also data storage (radial profiles occupy significantly less memory than 2-D $y$-stamps). Ideally, stacking (over galaxies and/or \textit{cells}) and azimuthal averaging (the process of extracting 1-D radial profiles from 2-D maps) are both linear processes. Thus, the final result should be similar irrespective of their order of execution. 

To extract the projected radial profiles of Compton-$y$, instead of circular apertures used in most previous studies, we implement annular apertures (same as the differential profiles of \citetalias{Das2023a}) because these are more sensitive to identifying any local deviation in thermal pressure from standard expectations. For the $cell_{ij}$ of halo mass ${\rm M}_{200,i}$ and redshift $z_j$, $y_{avg}(\theta_{ij})$ in Equation\,\ref{eq:ysub} and Equation\,\ref{eq:stack} is the azimuthally averaged Compton-$y$, i.e., average Compton-$y$ in an annular aperture of angular size $\theta_{ij}$ (Equation\,\ref{eq:f200}), 

\begin{equation}\label{eq:ysub}
\vspace{-0.1 in}
    y_{avg}(\theta_{ij}) = \Bar{y}(\theta_{ij}-\delta\theta\leqslant\theta<\theta_{ij}+\delta\theta)\\ %- \Bar{y}(\theta_{ij}+\delta\theta\leqslant\theta<\theta_{bkg})\\
    %{\rm where}\; \theta_{bkg} = \sqrt{ 2(\theta_{ij}+\delta\theta)^2 - (\theta_{ij}-\delta\theta)^2} \\
\end{equation}
\begin{equation}\label{eq:f200}
\begin{split}
     {\rm where}\; \theta_{ij} = r_{\perp,i}/{\rm D}_{{\rm A},j}; \;\;    r_{\perp,i} = f_{200}{\rm R}_{200,i} 
     \\
     {\rm and}\; \delta\theta = \delta f_{200} {\rm R}_{200,i}/{\rm D}_{{\rm A},j}
    \end{split}
    \vspace*{0.3 in}
\end{equation}
%and subtract $y_{null}$ from the actual $y$-stamp (Equation\,\ref{eq:ysub}).

$y_{avg}(\theta_{ij})$ is extracted at a fixed fraction $f_{200}$ of virial radius (Equation\,\ref{eq:f200}), as this provides a more accurate and easier-to-interpret result than that stacked at fixed angular size or fixed projected distance, especially if the sample spans a large range of redshift and virial mass \citepalias[see the appendix\,A of][for more details]{Das2023a}. 

\textbf{Step 8.} For each galaxy, we sum over all redshift \textit{slices} and relevant halo mass \textit{slices} weighted by their normalized probabilities to be within a particular 1) halo mass \textit{slice}, and 2) redshift \textit{slice}. Then we sum over galaxies weighted by their normalized probabilities of being star-forming or quiescent (Equation\,\ref{eq:stack}). 

We repeat the same stacking procedure separately for star-forming and quiescent galaxies.

\begin{widetext}
\begin{equation}\label{eq:stack}
\begin{split}
    y_{\rm stacked}(f_{200}) =
    \sum_{k} w_{\scriptsize{{\rm SF},k}} \sum_i w({\rm M}_{200{\rm-SF},i}) \sum_j  w(z_j) y_{avg}(\theta_{ij} | z_j,{\rm M}_{200,i}) 
    {\;\;\rm for\;starforming\;galaxies,\;and} \\
    = 
    \sum_{k} w_{\scriptsize{{\rm Q},k}} \sum_i w({\rm M}_{200{\rm-Q},i}) \sum_j  w(z_j) y_{avg}(\theta_{ij} | z_j,{\rm M}_{200,i}) 
    {\;\;\rm for\;quiescent\;galaxies} 
\end{split}
\end{equation}     
\end{widetext}

\textbf{Step 9.} Finally, we use the Bootstrap method to estimate the unbiased mean and uncertainties of $y_{\rm stacked}$ that will be used in \S\ref{sec:P} and the following sections. For a sample of N galaxies, we create a replica of $y_{\rm stacked}$ by replacing one galaxy with a randomly chosen galaxy from the rest N-1 galaxies in that sample. For every stack, we make a set of 1000 replicas and obtain the mean, median, and covariance matrix \textbf{C$_{\rm stacked}$} of those replicas. We compare the median and mean to confirm that the Compton-$y$ distribution across galaxies is not skewed. Note that \textbf{C$_{\rm stacked}$} includes statistical as well as systematic uncertainties, and because of the large sample size, it is dominated by systematic uncertainties. 

%\footnote{This is more conservative than Bootstrap; its estimated uncertainties are larger than Bootstrap, and it is reproducible, unlike Bootstrap. We compare the median and mean to confirm that the distribution of Compton-$y$ is not skewed; thus the computationally expensive Bootstrap is not required}. 

%To assess the systematics due to intrinsic differences among individual galaxies, we construct 10 subsamples each 1/10th of the sample size by randomly drawing galaxies from the sample weighted by the normalized probability and then repeating the above analysis for each subsample. 
\end{subequations}

%We estimate the uncertainties in stacked Compton-$y$ value using the bootstrap method. For a sample of N galaxies, we create a replica of the sample by replacing one galaxy with a randomly chosen galaxy from the rest N-1 galaxies in that sample. For every stack, we make a set of 1000 replicas and obtain the mean and covariance matrix of those replicas.

% redshift bin 0.02 dex, virial mass bin 0.1 dex. Separately done for SF and Q. Same galaxy with different w_SF, M200, and w(M200).  
% Instead of stacking a 2d map that requires reconstruction we extract a 1 d profile and then stack. Faster, less assumption.
%only differential
%battaglia2012 profile add. 

%adaptive mass bin. in virial mass separately for star-forming and quiescent 
%adaptive stamp size so that at least 2 Mpc is probed.
%cubic interpolation
%>50% NAN ignored
%bootstrap replica 1000 or 1% of the sample size whichever is larger. 
%variable zero point handled. cumulative main. differential for cross-checking. 

\section{Other mass bins}
\begin{figure*}
    \centering
    \includegraphics[width=0.49\linewidth]{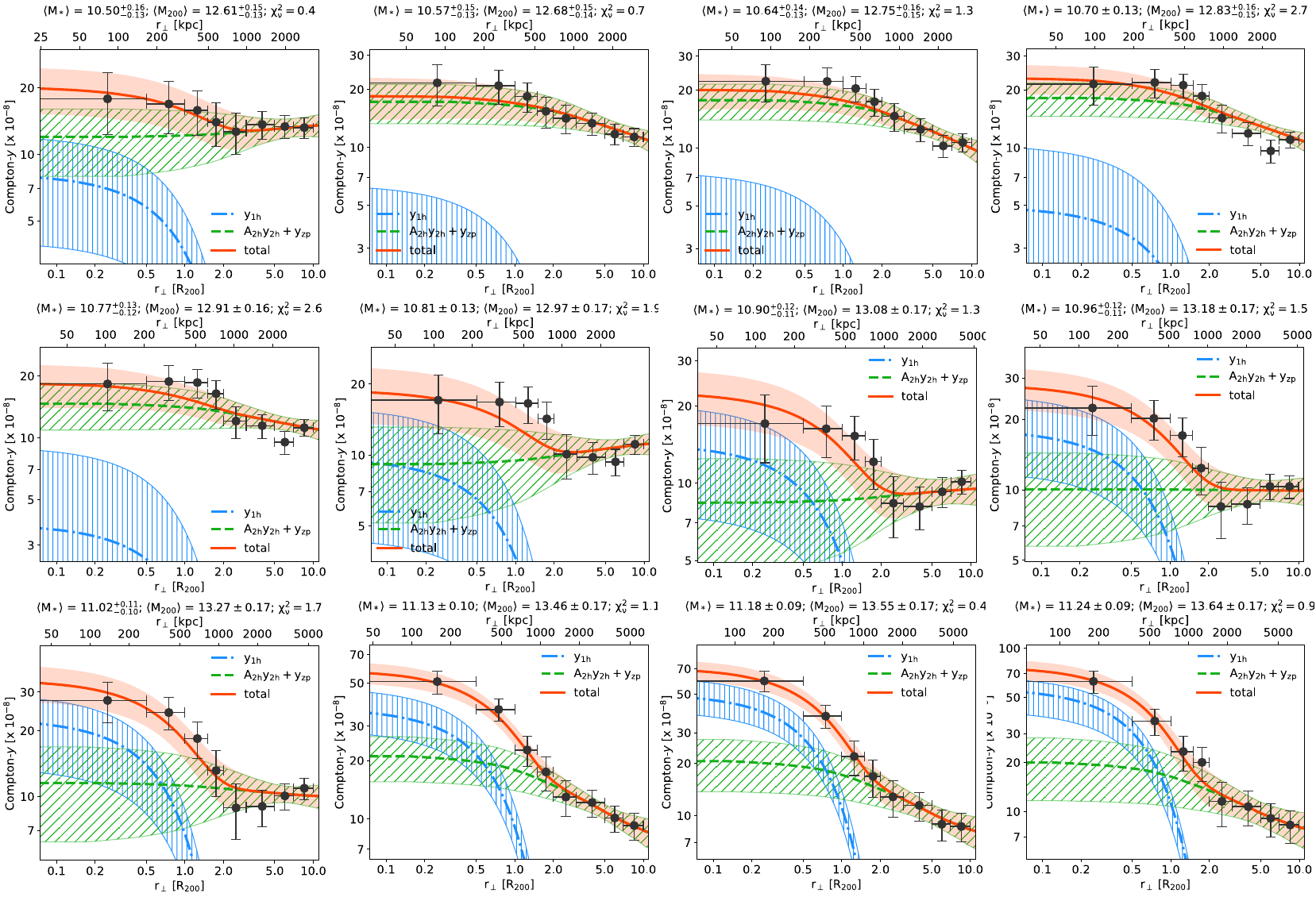}
    \includegraphics[width=0.49\linewidth]{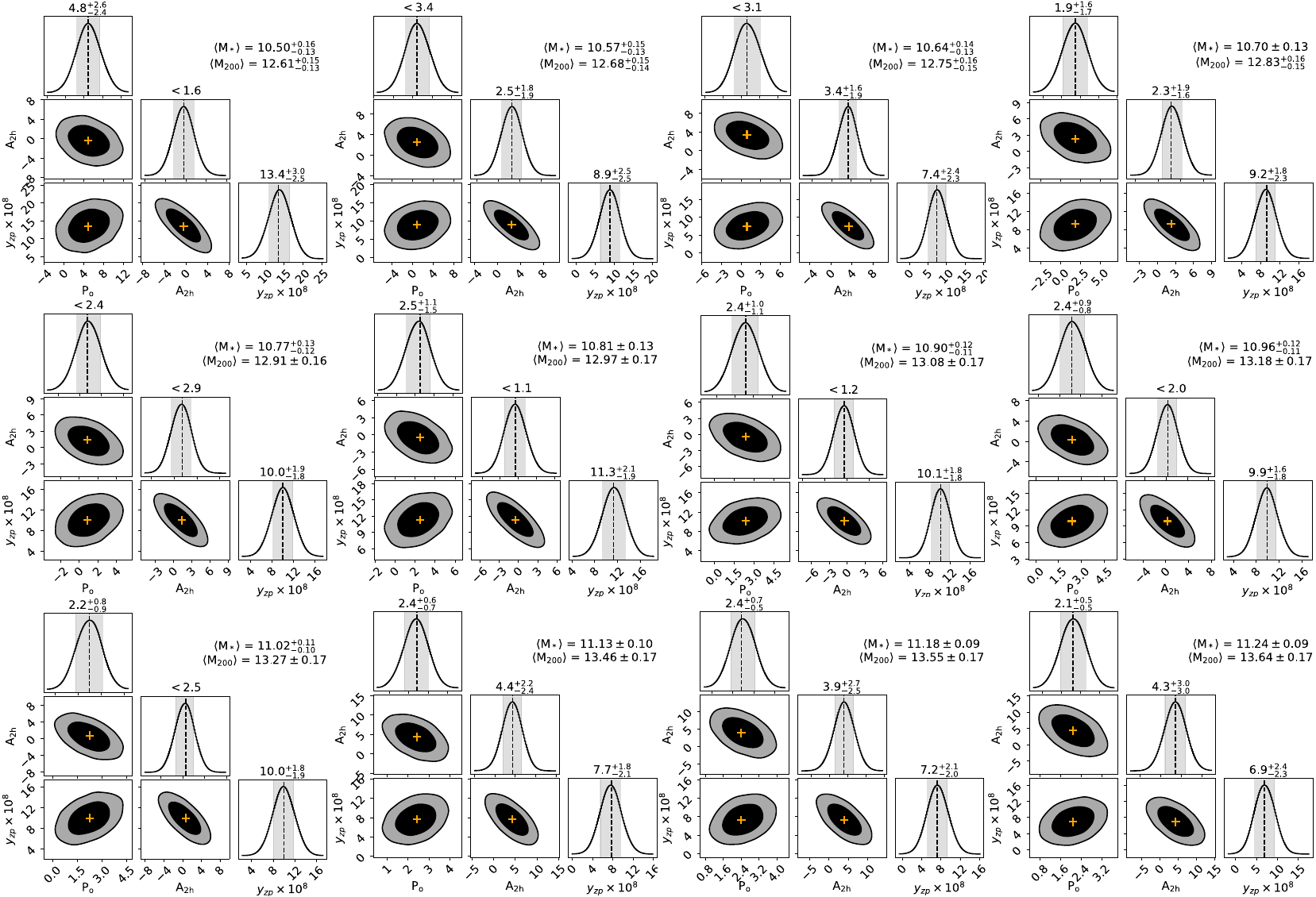}
    \includegraphics[width=0.49\linewidth]{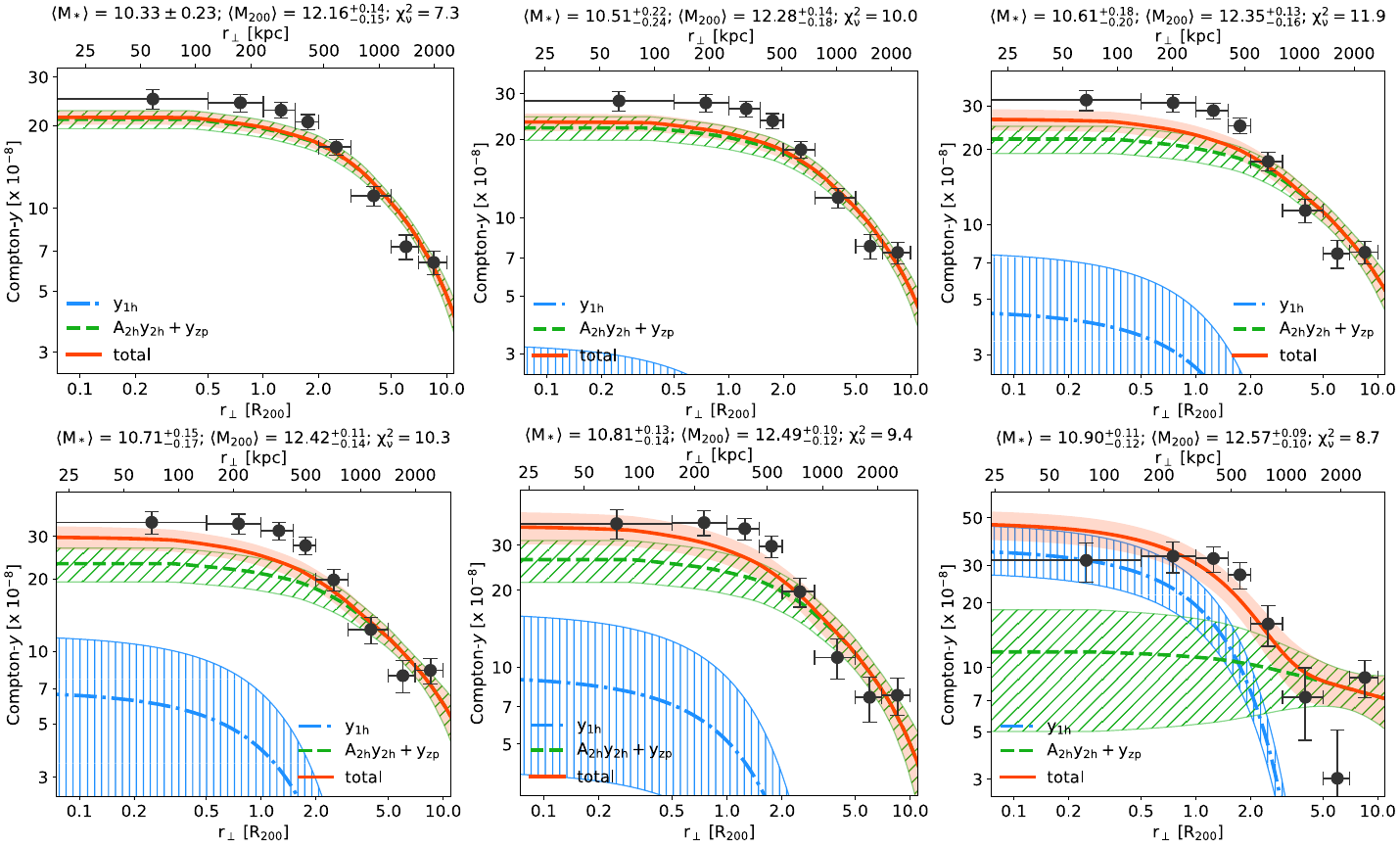}
    \includegraphics[width=0.49\linewidth]{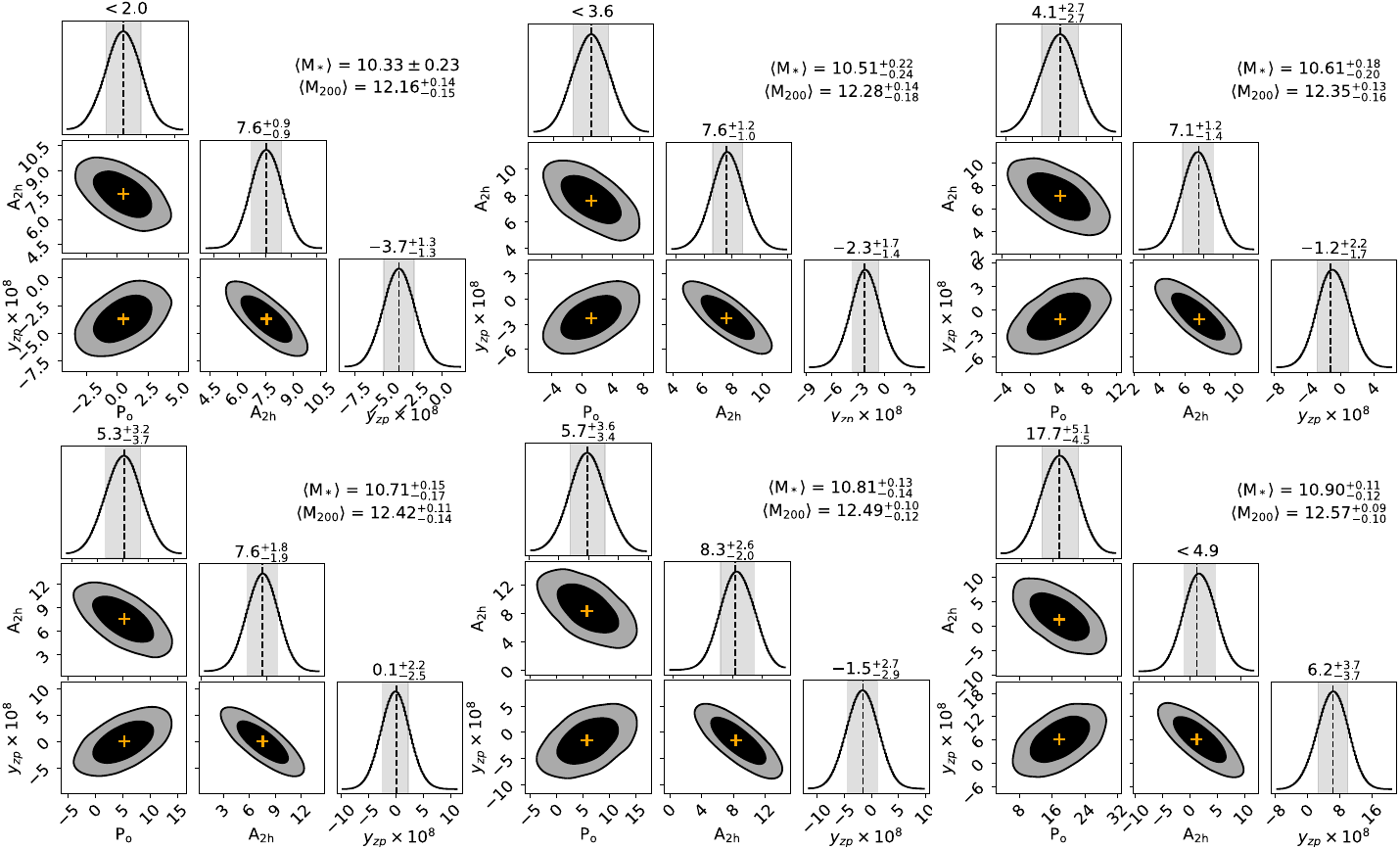}
    \caption{Top three rows: Quiescent galaxies. Bottom two rows: Star-forming galaxies. See Figure\,\ref{fig:bestfit} for the description.}
    \label{fig:all}
\end{figure*}

The projected radial profile of stacked Compton-$y$ with the best fitted model, the covariance matrices, and the posterior distribution of the fitting parameters for different quiescent and star-forming galaxies are shown in Figure\,\ref{fig:all}.

\textcolor{black}{Quiescent galaxies are generally better fitted than star-forming galaxies because of their stronger ``1-halo" term, larger size, and weaker ``2-halo" term than star-forming galaxies. Additionally, due to larger sample sizes and possibly smaller galaxy-to-galaxy scatter, the covariance matrices have smaller values in star-forming galaxies compared to quiescent galaxies. This pushes the $\chi^2$ to higher values in star-forming galaxies for comparable offset of the model from the data in both galaxy populations. Testing the \textit{true} shape of the pressure profile in star-forming galaxies would be feasible with next-generation mm data featuring better angular resolution.}

\section{Could the pressure profile of star-forming galaxies be flatter?}\label{sec:flat}

\begin{figure*}
\includegraphics[width=0.475\textwidth]{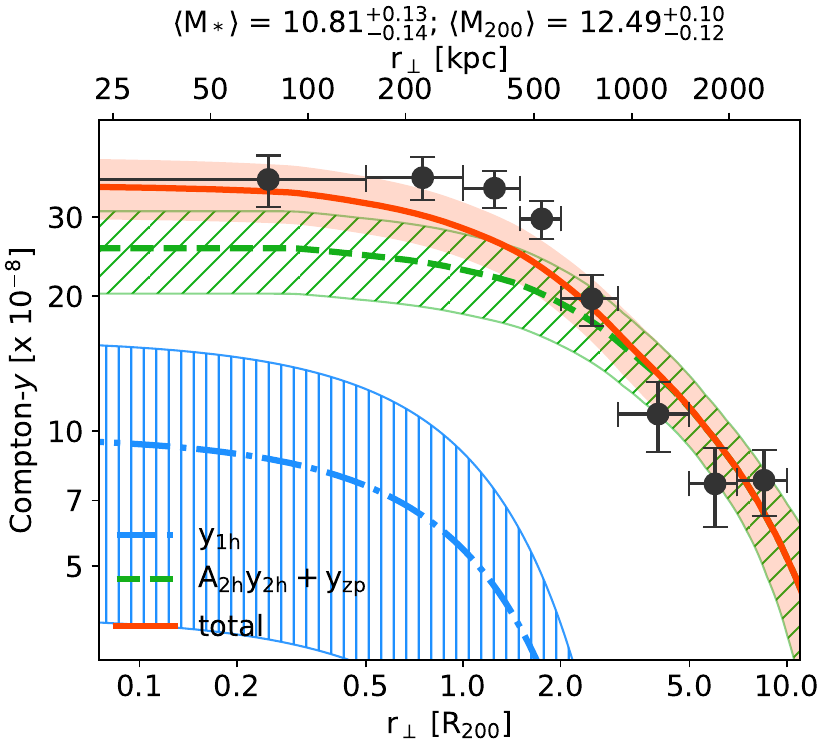}
\includegraphics[width=0.475\textwidth]{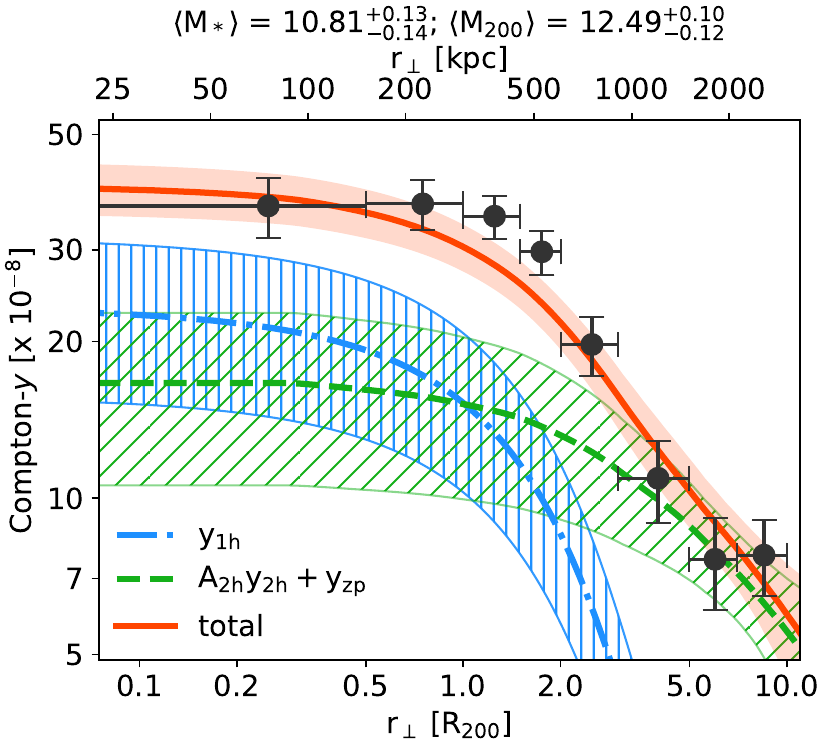}
\includegraphics[width=0.475\textwidth]{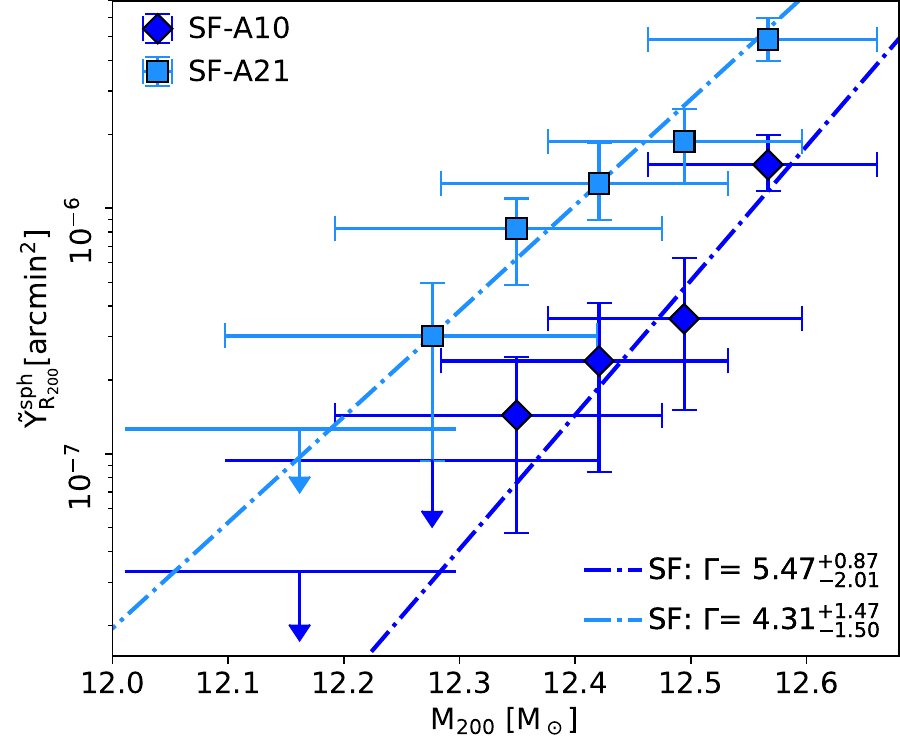}
\includegraphics[width=0.475\textwidth]{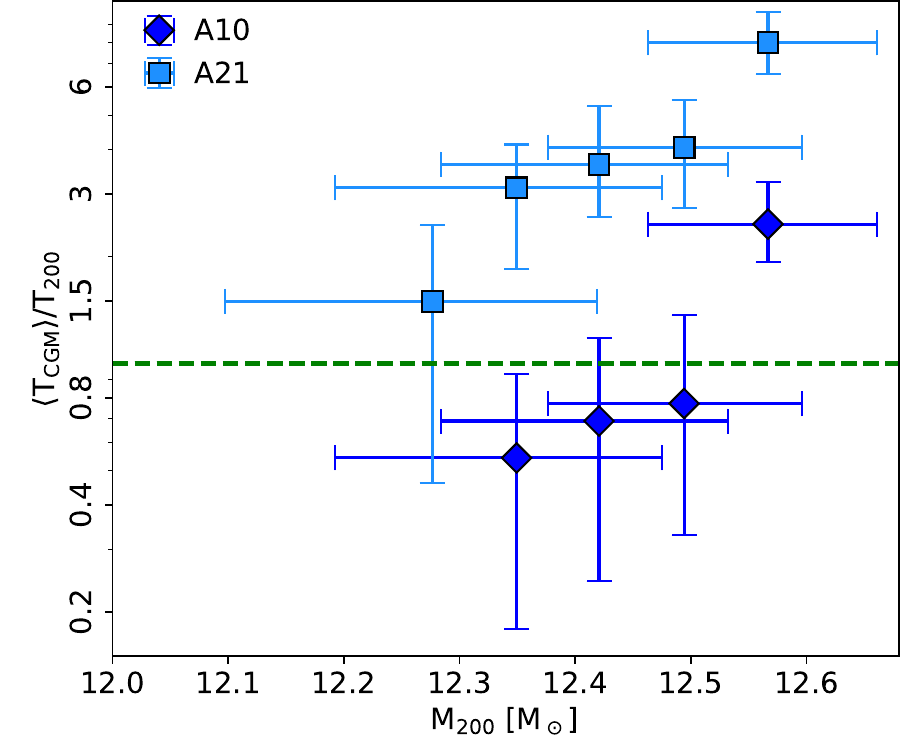}
\caption{Comparison between standard \citepalias{Arnaud2010} and flatter \citepalias{Amodeo2021} GNFW pressure profile for the ``1-halo" term. Top: The best-fitted model for one of the mass bins: left- \citetalias{Arnaud2010}, right- \citetalias{Amodeo2021}. See Figure\,\ref{fig:bestfit}, \ref{fig:YvsM}, and \ref{fig:fb} for the description.}\label{fig:flat}
\end{figure*}

To test if a pressure profile flatter than standard GNFW describes the data for star-forming galaxies better, we repeat the main analysis with the shape of ``1-halo" term replaced by the best-fit results of \citeauthor{Amodeo2021} (\citeyear[hereafter \citetalias{Amodeo2021}]{Amodeo2021}): $\alpha = 0.8, \beta = 2.6$ in Equation\,\ref{eq:pgnfw}. For convenience, we will refer to this model as ``flatter GNFW" and the model of \citeauthor{Arnaud2010} (\citeyear[hereafter \citetalias{Arnaud2010}]{Arnaud2010}) as ``standard GNFW". Statistically, flatter profile is better (AIC\footnote{Akaike Information Criterion ${\rm AIC} = 2n - \rm ln(\hat L)$, $n$ no. of fitted parameters, $\rm ln(\hat L)$ natural log of likelihood, see equation\,\ref{eq:chisq}} smaller, $\rm AIC_{standard} - AIC_{flatter} >2$) for all halos except the smallest mass bin. For both the profiles, $\Gamma>5/3$ (Figure \ref{fig:flat}, bottom left). $\Gamma>5/3$ would result from the increasing effect of stellar feedback in a flatter profile and metal line-driven cooling in the standard GNFW profile (see \S\ref{sec:scenario}). Despite being a better statistical fit, the flatter profile leads to unphysical results because of larger \textit{normalization} than standard GNFW (Figure \ref{fig:flat}, top panels). The halos become overabundant in baryons ($f_b>f_{b,cosmo}$) if they are at virial temperature. Or, the volume-average temperature has to be super-virial at $3.6_{-0.6}^{+0.7}\rm T_{200}$ (Figure \ref{fig:flat}, bottom right) if they satisfy baryon sufficiency, moving the entire halo out of thermal equilibrium. Therefore, standard GNFW is preferred on physical ground in $\rm M_{200} \geq 10^{12.3}$\msun halos. In $\rm M_{200} < 10^{12.3}$\msun halos, the ``1-halo" term cannot be constrained anyway; the standard and flatter profiles are statistically and physically indistinguishable because of the dominant tSZ background. 

We clarify that while we rule out the possibility of volume-average super-virial temperature, the existence of super-virial CGM is not impossible. It has been detected in the Milky Way both in X-ray absorption \citep{Das2019a,Das2021b,Lara2023a,McClain2024,Lara2024} and in X-ray emission \citep{Das2019c,Gupta2021,Bluem2022,Bhatt2023,Ponti2023,Gupta2023} coexisting with the virial phase across the sky. However, the super-virial phase is traced by highly ionized metals in X-ray ($\alpha$ elements in absorption and Fe L shells in emission). Therefore, it primarily contributes to the metal budget, and its contribution to the baryon budget is subject to the assumption of the absolute metallicity. Secondly, the super-virial emitting phase likely occupies the inner CGM \citep{Bisht2024}, so its contribution to the overall baryon budget is not significant. The super-virial absorbing phase is less dense and extended out to a large radius, but it does not have any observational evidence to be volume-filling. Therefore, its contribution to the tSZ effect would be relatively small. Thus, while the CGM of star-forming galaxies in our sample could have a super-virial phase, its effect would not be reflected in the volume-average estimates of thermal energy or baryon fraction integrated over \rvirialn, i.e., the entire CGM. 

%\bibliography{reference,reference_2}{}
\bibliographystyle{aasjournal}

\end{document}